\documentclass[%
 reprint,
 amsmath,amssymb,
 aps,
]{revtex4-2}

\usepackage{graphicx}
\usepackage{dcolumn}
\usepackage{bm}

\begin{document}

\preprint{APS/123-QED}

\title{On payload architecture and pointing control strategies for TianQin}

\author{Yuzhou Fang}
\author{Xuefeng Zhang}
 \email{zhangxf38@sysu.edu.cn}
\author{Fangyuan Fu}
\author{Hongyin Li}
\affiliation{MOE Key Laboratory of TianQin Mission, TianQin Research Center for Gravitational Physics $\&$ School of Physics and Astronomy, Frontiers Science Center for TianQin, Gravitational Wave Research Center of CNSA, Sun Yat-sen University (Zhuhai Campus), Zhuhai 519082, China.}

\date{\today}

\begin{abstract}
TianQin is a proposed mission for space-based gravitational-wave detection that features a triangular constellation in circular high Earth orbits. The mission entails three drag-free controlled satellites and long-range laser interferometry with stringent beam pointing requirements at remote satellites. For the payload architecture and pointing control strategies, having two test masses per satellite, one for each laser arm, and rotating entire opto-mechanical assemblies (each consisting of a telescope, an optical bench, an inertial sensor, etc.) for constellation breathing angle compensation represent an important option for TianQin. In this paper, we examine its applicability from the perspectives of test mass and satellite control in the science mode, taking into account of perturbed orbits and orbital gravity gradients. First, based on the orbit-attitude coupling relationship, the required electrostatic forces and torques for the test mass suspension control are estimated and found to be sufficiently small for the acceleration noise budget. Further optimization favors configuring the centers of masses of the two test masses collinear and equidistant with the center of mass of the satellite, and slightly offsetting the assembly pivots from the electrode housing centers forward along the sensitive axes. Second, the required control forces and torques on the satellites are calculated, and thrust allocation solutions are found under the constraint of having a flat-top sunshield on the satellite with varying solar angles. The findings give a green light to adopting the two test masses and telescope pointing scheme for TianQin. 
\end{abstract}

\maketitle


\section {Introduction}\label{sec:intro}
\begin{figure}[ht]
\centering
\includegraphics[width=0.48\textwidth]{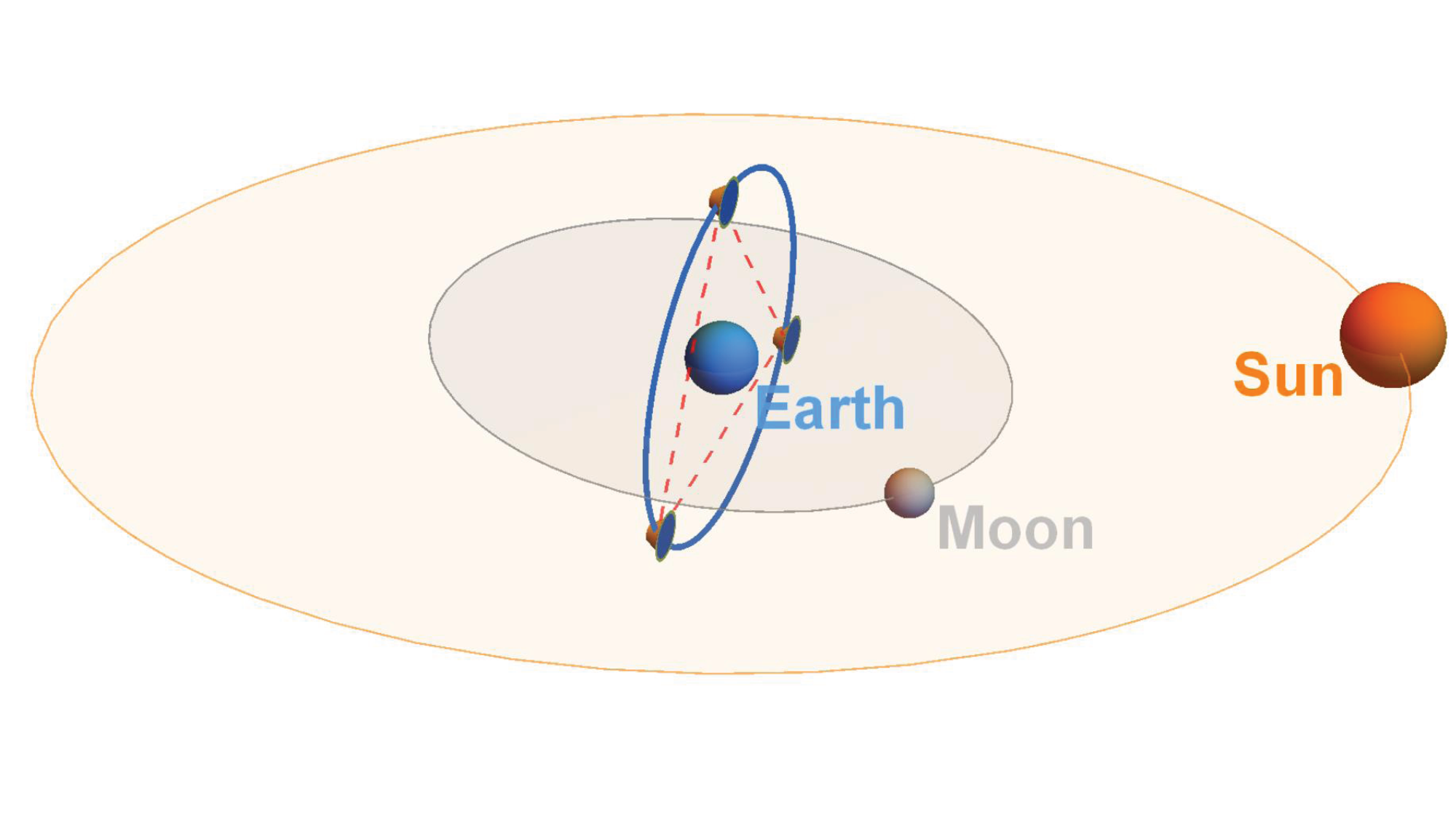}
\caption{\label{fig:EarthMoonSun} Depiction of the TianQin constellation revolving around the Earth and subject to varying incoming sunlight (not to scale) \cite{ye2021eclipse}. }
\end{figure}

The TianQin mission plans to deploy three drag-free controlled satellites in a nearly equilateral-triangle constellation with arm lengths of $\sim1.7\times 10^5$ km. The satellites follow circular orbits around the Earth at an altitude of $\sim10^5$ km (see Fig. \ref{fig:EarthMoonSun}), and the constellation plane is facing a verification source, the white-dwarf binary RX J0806.3+1527 \cite{luo2016TianQin}. The mission is to detect gravitational wave (GW) signals in the frequency range of $10^{-4}$ Hz to 1 Hz, and the sensitivity goal presents great challenges to the science payloads and satellite platforms. 

Space-based detectors differ from ground-based detectors in many important aspects. Perhaps most prominently, space-based detectors generally do not have laser armlengths and beam pointings fixed due to orbital dynamics. Though the variations should be kept as small as possible by orbit and constellation design, eliminating them is generally not practical. Taking TianQin for example, the estimated deviation from an ideal equilateral triangle due to lunisolar gravitational perturbation is 0.1\% in the armlengths and $60\pm0.1^\circ$ in the breathing angles with a typical period of 3.6 days \cite{ye2019optimizing}. Additionally, these is slight wobbling of the constellation plane ($\sim 0.05^\circ$ per 3.6 days) mostly due to the Moon's gravitational pull from sideways. These basic operating condition and environment for space-based detectors have profound influence on measurement principles, science payload design, satellite control, and data processing on ground, and must be dealt with on the mission and system levels from the very beginning. 

The key science payload of space-based detectors like LISA \cite{sallusti2009lisa, lisa2017proposal, danzmann2011lisa} and TianQin mainly consists of inertial sensors, enclosing free-falling test masses (TMs) inside as reference end mirrors, and long-range laser interferometers for measuring tiny distance changes between TMs caused by GWs. Given an operating environment, the various ways the key payload is configured from basic components to achieve high precision TM-to-TM measurements are referred to as payload architectures in this paper. Closely interrelated to payload architectures are the payload control and operation \cite{gath2007drag}. For nominal science observation, one crucial aspect of the payload control is the pointing of the outgoing laser beams at distant satellites. The pm/Hz$^{1/2}$-level interferometric measurement demands a pointing requirement of $\sim 10$ nrad in DC bias and $\sim 10$ nrad/Hz$^{1/2}$ in jitters \cite{lisa2017proposal}. This poses a great challenge to the fine pointing control and its design. 

\begin{figure}[htb]
\centering
\includegraphics[width=0.48\textwidth]{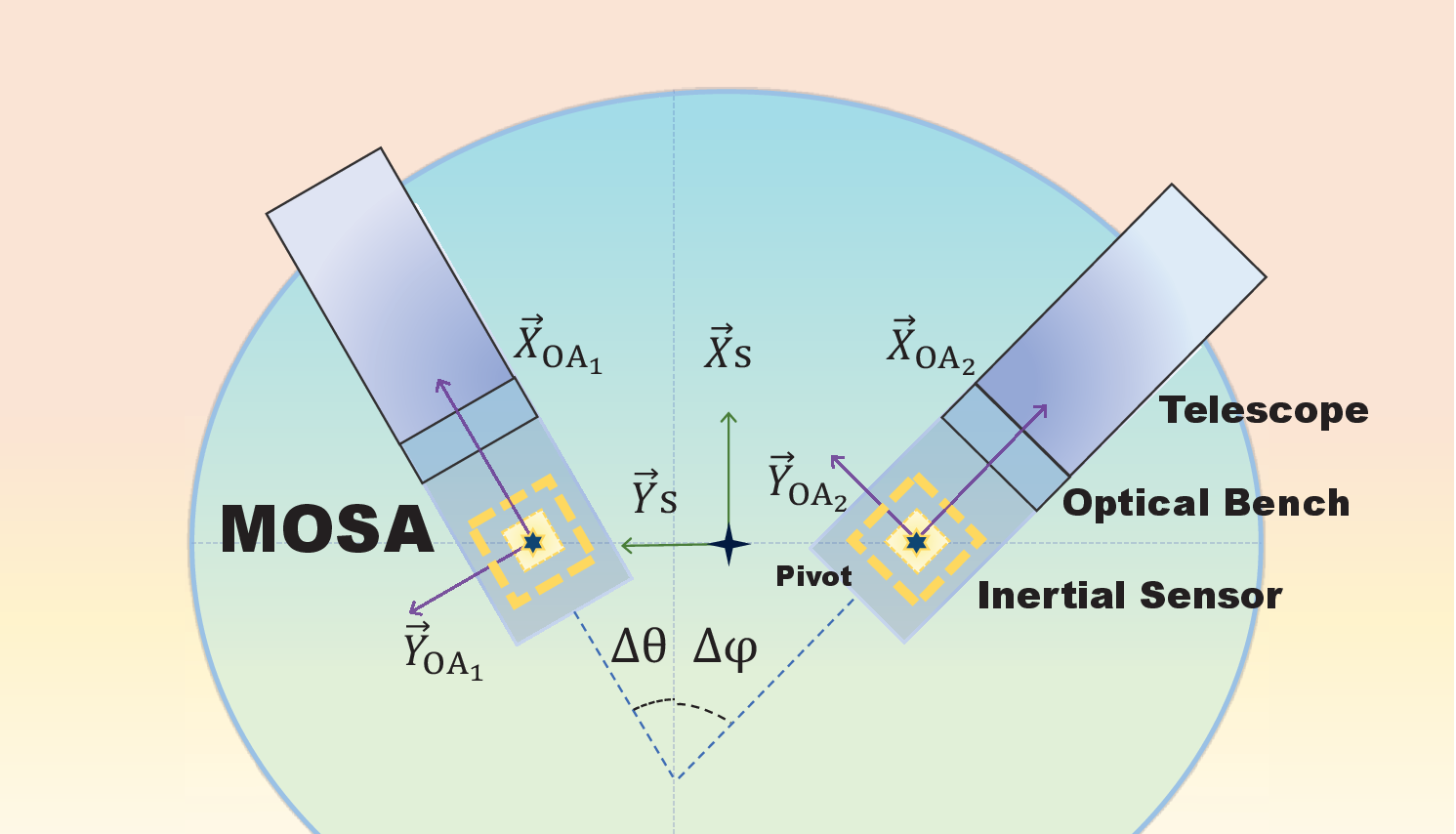}
\caption{\label{fig:SC} Illustration of Movable Optical
Sub-Assemblies (MOSAs). The axes depict the body-fixed coordinate systems of the satellites ($\vec{X}_\mathrm{S}$, $\vec{Y}_\mathrm{S}$) and two MOSAs ($\vec{X}_{\mathrm{OA}_{1,2}}$, $\vec{Y}_{\mathrm{OA}_{1,2}}$). The MOSAs can rotate about the $\vec{Z}_{\mathrm{OA}_{1,2}}$-axes, respectively. }
\end{figure}

\begin{figure}[htb]
\centering
\includegraphics[width=0.48\textwidth]{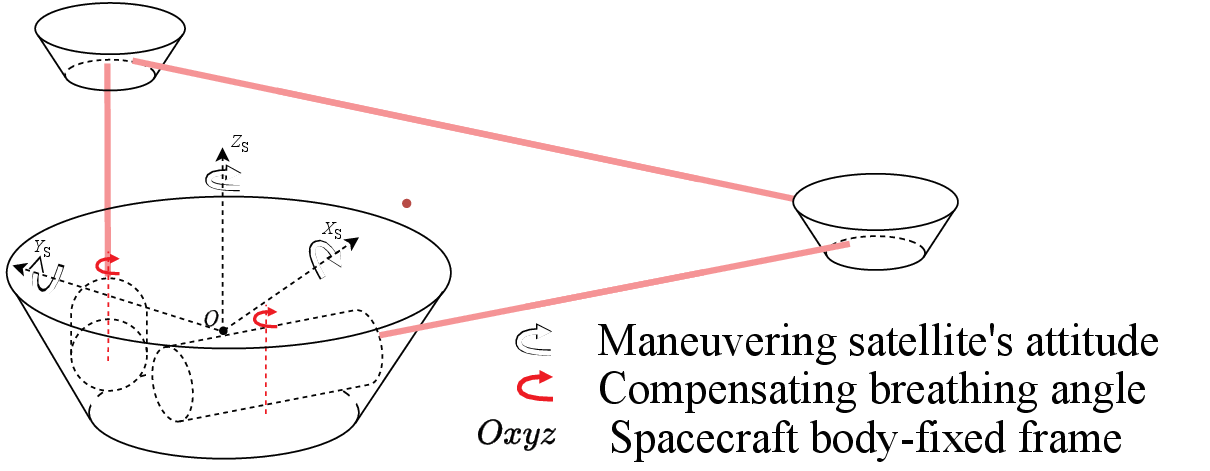}
\caption{\label{fig:Formation} The telescope pointing scheme relies on the MOSA articulation to compensate for breathing angle variations within the constellation plane, and the satellite attitude control to align the laser beams in off-plane directions.}
\end{figure}

To tackle special needs of space-based GW detectors, several options for payload architectures and pointing control strategies have been proposed and studied in literature (see, e.g., \cite{gath2006lisa, gath2009lisa, gath2010challenges}). LISA, as the pioneer in space-based GW detection, has opted for the design baseline that each spacecraft are to be equipped with two Movable Optical Sub-Assemblies (MOSAs) as shown in Fig. \ref{fig:SC} \cite{lisa2017proposal, weise2017opto}. The assembly consists of a telescope, an optical bench, an inertial sensor (also known as gravitational reference sensor), supporting structures, and pivot mechanisms. The two cubic TMs are allowed to free-float in the directions of the laser arms and suspended by electrostatic forces in the other degrees of freedom to maintain nominal TM position and alignment inside the electrode housings (EHs). To account for the annual variation ($\pm 1^\circ$) of the breathing angles, the entire MOSAs can be rotated about the pivot axes vertical to the constellation plane (see Fig. \ref{fig:Formation}). Meanwhile, pointing adjustment of the outgoing laser beams in off-plane directions is carried out by the drag-free and attitude control (DFAC) of the spacecraft with the help of micro-Newton thrusters \cite{LISA2000}. The whole design is dubbed the two-TM and telescope pointing scheme. 

A competing option considered in LISA's trade-off studies \cite{johann2008european, lisa2017proposal} is to have one inertial sensor/TM, one common optical bench, and two telescopes, all rigidly fixed to one another and to one spacecraft. To correct for constellation breathing, the telescope is designed to have a wide field of view, and the outgoing beam direction relative to the telescope axis can be adjusted by actuating steering mirrors on the optical bench \cite{johann2006novel, weise2009alternative, brugger2017experiment}. TM shapes can take on multiple forms, allowing cubic, quasi-cubic, and spherical alternatives \cite{gerardi2014invited, liu2022single}. The entire design is dubbed the single-TM and in-field pointing scheme. 

Both schemes have their own pros and cons \cite{gerardi2014invited}. For example, a prominent obstacle with the in-field pointing is the tilt-to-length coupling (see, e.g., \cite{hasselmann2021lisa, liu2022single}), which is made even more challenging by the telescope magnification ($\sim 150$). Nevertheless, having two TMs per spacecraft and articulating the MOSAs render the payload and spacecraft control quite complicated. For TianQin, the in-field pointing appears intriguing, given that TianQin has relatively small variations ($\pm 0.1^\circ$) in breathing angles. However, the feasibility studies are still on-going. 

The geocentric design of TianQin has taken into account engineering benefits in satellite deployment, orbit determination, data communication, etc. Since its earlier conceptualization, questions have been raised regarding its payload architecture and pointing control strategy, and particularly, how to make the overall design well suited to the geocentric orbits. Apparently, the two-TM and telescope pointing scheme presents an important candidate. Nevertheless, without in-depth modeling and assessment, it is not certain whether the scheme can be applied to TianQin and whether certain modification is needed. Hence, the paper is intended to address this fundamental issue of the TianQin mission, and pave the way for future system development. For related studies on TianQin DFAC, one may refer to, e.g. \cite{lian2021determination, deng2022frequency, xiao2022drag, ma2022controller, wang2023stacked, zhang2023DYN}. Most of these studies focused on developing control algorithms, and the joint dynamics and control of the MOSAs and TMs are yet to be included. 

The applicability issue can be examined from two main perspectives, both based on numerical orbits and the orbit-attitude coupling derived from the telescope pointing scheme (Sec. \ref{sec:modelsetup}). First, for the TM control, we calculate the required electrostatic forces and torques on the TMs and compare them with the allowed maximum values derived from the acceleration noise budget (Sec. \ref{sec:SuspensionC}). Second, for the satellite control, we calculate the required forces and torques on the satellites, and show that they can be fulfilled by the micro-propulsion system under the design constraints of the satellites (Sec. \ref{sec:AttitudeofSatellite}).

\section{Model Setup}\label{sec:modelsetup}
This section elaborates on the dynamic models used in this study, including initial orbital parameters, coordinate systems, orbit-attitude coupling, and a satellite model. The orbit-attitude coupling relationship is a direct consequence of the pointing control strategy. It plays a key role in our modeling and assessment that can dispenses with detailed control algorithms. 

\subsection{Numerical orbits}\label{sec:orbit}
Calculating the nominal attitudes of the satellites and MOSAs relies on having TianQin's orbit information during the mission. We have used General Mission Analysis Tool (\texttt{GMAT}) \cite{hughes2014GMATveri} software for orbit simulation, and obtained numerical data of the satellites' positions, velocities, accelerations, and gravity gradients. Propagating realistic perturbed orbits involves various force models, which include a 10$\times$10 spherical harmonic representation of the Earth's gravity field (JGM-3 \cite{tapley1996joint}), a point-mass model for the Sun, the Moon, and other planets in the solar system (the ephemeris DE421 \cite{folkner2009planetary}), and the first-order relativistic effect. 

\begin{figure}[ht]
  \centering
  \includegraphics[width=0.48\textwidth]{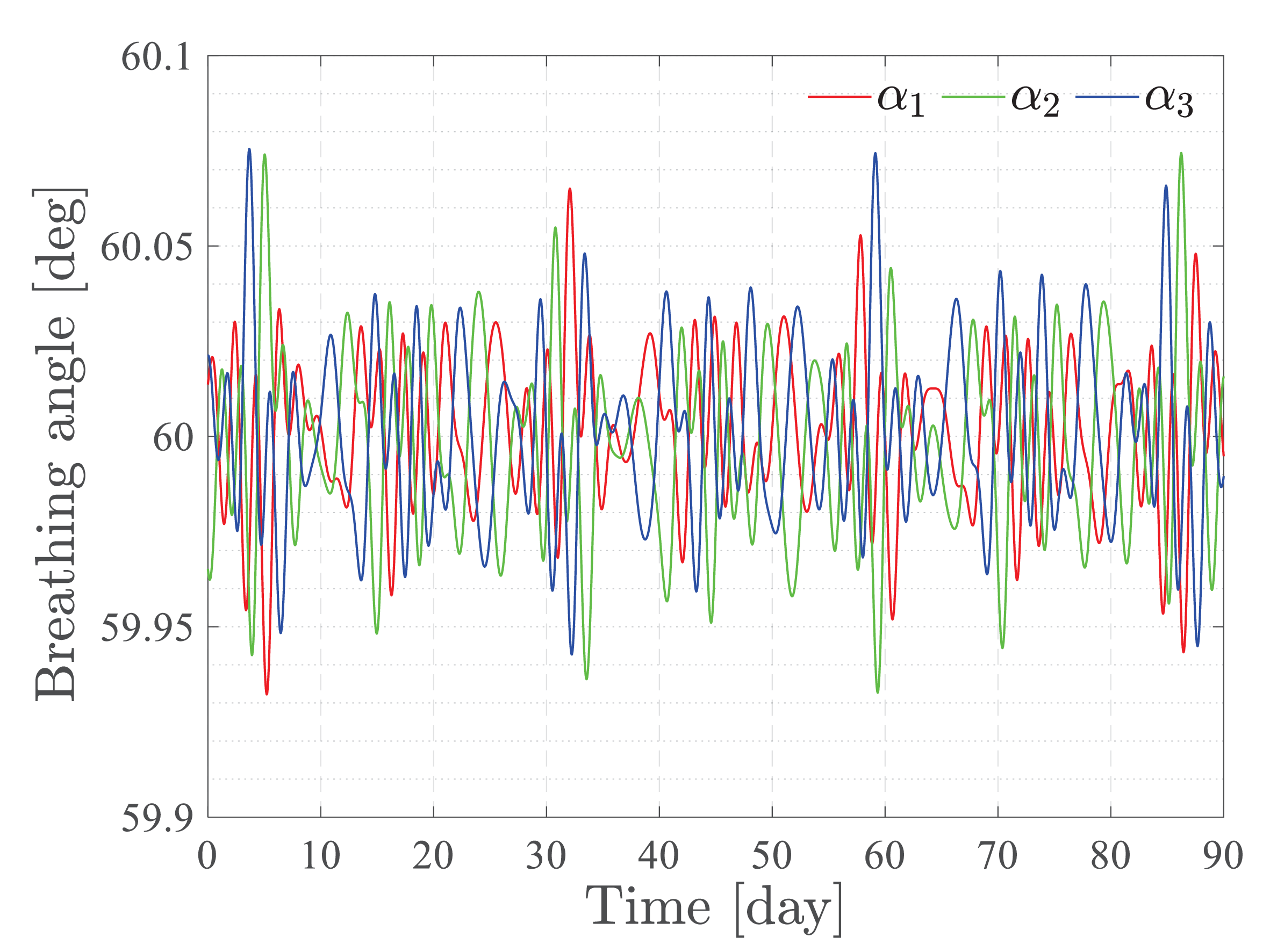}
\caption{\label{fig:breathAngle} Time variations of the three breathing angles $\alpha_{1,2,3}$ of the constellation. }
\end{figure}

\begin{figure}[ht]
  \centering
  \includegraphics[width=0.48\textwidth]{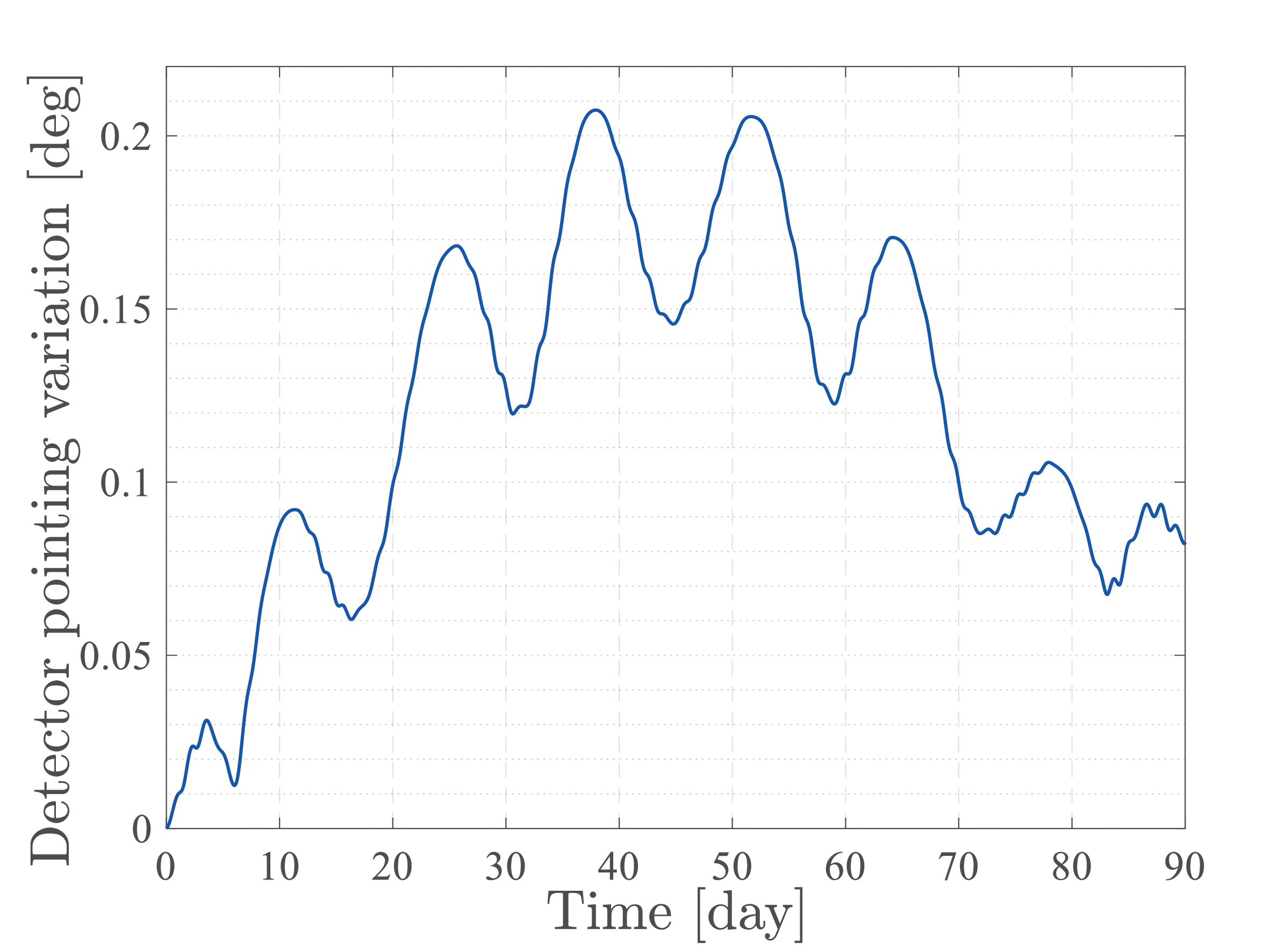}
\caption{\label{fig:pointingVariation} Angle variation of the normal of the constellation plane from its initial direction. }
\end{figure}

Since the satellites are drag-free controlled with high precision, we assume that the orbital evolution of the center of mass (CoM) of the satellite is under pure gravity and that the coupling with DFAC is currently neglected (see the Appendix \ref{App:orbit_dev} for estimated deviation from pure gravity orbits). The initial orbital elements are from our previous research \cite{ye2021eclipse}, and listed in Table \ref{tab:OP}. They have been optimized to satisfy the configuration stability criteria for the TianQin constellation. More relevant to the pointing control, the time evolution of the breathing angles are illustrated in Fig. \ref{fig:breathAngle}, and the angle variation of the normal of the constellation plane (detector pointing) from its initial direction is given in Fig. \ref{fig:pointingVariation}. 

\begin{table}[ht]
\caption{\label{tab:OP}The optimized initial orbital elements for three TianQin satellites (SC1, 2, 3) in the J2000-based Earth-centered ecliptic coordinates at the epoch 22 May, 2034 12:00:00 UTC \cite{ye2021eclipse}. They can be easily converted to equatorial coordinates. The notations used below represent various orbital elements: $a$ for the semi-major axis, $e$ for the eccentricity, $i$ for the inclination, $\Omega$ for the longitude of the ascending node, $\omega$ for the argument of periapsis, and $\nu$ for the true anomaly \cite{orbitElement}.}
\begin{ruledtabular}
\begin{tabular}{ccccccc}
 & $ a $\,(km) &  $ e $ & $ i $\,($ ^{\circ} $) \\ 
\hline
SC1 & 100\,926.158\,459 & 0.000\,300 & 94.774\,822  \\
SC2 & 100\,940.789\,023 & 0.000\,019 & 94.782\,183  \\
SC3 & 100\,938.056\,412 & 0.000\,411 & 94.785\,623  \\
\hline
\hline
 & $ \Omega $\,($ ^{\circ} $) & $ \omega $\,($ ^{\circ} $) & $ \nu $\,($ ^{\circ} $) \\
\hline
SC1 & 209.433\,009 & \hphantom{0}\hphantom{0}0.980\,870 & \hphantom{0}84.729\,131 \\
SC2 & 209.430\,454 & 205.692\,143 & 359.976\,125 \\
SC3 & 209.438\,226 & \hphantom{0}\hphantom{0}0.061\,831 & 325.619\,846 \\
\end{tabular}
\end{ruledtabular}
\end{table}

\subsection{Coordinate systems}\label{sec:coordinate}
The dynamic model capturing three satellites, each with two MOSAs and two TMs, invokes multiple reference frames (see Fig. \ref{fig:SC}). 

\subsubsection{Inertial and body-fixed reference frames}
The body-fixed frames, as each of them attaches to one existing body, are defined as follows. 

\begin{itemize}
   \item The $\mathcal{I}$ frame is the inertial frame to describe satellite motion in geocentric orbits. This article has used the J2000-based Earth-centered equatorial coordinate system with the axes $\vec{X}_\mathrm{I}$, $\vec{Y}_\mathrm{I}$, and $\vec{Z}_\mathrm{I}$. 

   \item The $\mathcal{S}_i$ frame ($i=1,2,3$) is rigidly attached to one satellite and describes its motion. It is built in the following. 
   \begin{itemize}
       \item The origin is at the CoM of the satellite. 
       \item $\vec{X}_{\mathrm{S}_i}$ bisects the $60^{\circ}$ angle between the two optical assemblies. 
       \item $\vec Z_{\mathrm{S}_i}$ is perpendicular to the solar panel. 
       \item $\vec Y_{\mathrm{S}_i}$ is determined by the right-hand rule. 
   \end{itemize}
   
   \item The $\mathcal{OA}_l$ frame ($l=1,2$) describes motion of one MOSA (also known as the optical assembly). 
   \begin{itemize}
       \item The origin is at the center of the electrode housing. 
       \item $\vec{X}_{\mathrm{OA}_l}$ aligns with the optical axis of the telescope. 
       \item $\vec{Z}_{\mathrm{OA}_l}$ aligns with $\vec{Z}_\mathrm{S}$
       \item $\vec{Y}_{\mathrm{OA}_l}$ is determined by the right-hand rule. 
   \end{itemize}
   
   \item The $\mathcal{TM}_l$ frame ($l=1,2$) describes motion of the TM. 
   \begin{itemize}
       \item The origin is at the CoM of the TM.
       \item The $x$, $z$, and $y$-axes are orthogonal to the TM faces and align with $\mathcal{OA}_l$ when the TM is in the nominal position. 
   \end{itemize}
\end{itemize}

\subsubsection{Target reference frames}
The three satellites need to be oriented with respect to one another. Their nominal attitudes are strictly dependent on the constellation, as the telescopes point at distant satellites (see Fig. \ref{fig:Formation}). Hence the target reference frames can be defined as follows. 

\begin{itemize}
   \item The $\mathcal{S}_i^*$ frame ($i=1,2,3$) describes the target/nominal attitude for one satellite. 
   \begin{itemize}
       \item The origin coincides with the nominal orbit of the satellite.
       \item $\vec X_{S_i^*}$ point towards the incenter of the triangular constellation. 
       \item $\vec Z_{S_i^*}$ is orthogonal to the constellation plane.
       \item $\vec Y_{S_i^*}$ is built from the cross product of the two above.
   \end{itemize}
   
   \item The $\mathcal{OA}_l^*$ frame ($l=1,2$) describes the target/nominal attitude of one optical assembly.
   \begin{itemize}
       \item The origin coincides with the one of $\mathcal{OA}_l$.
       \item $\vec{X}_{\mathrm{OA}_l^*}$ points at the distant satellite.  
       \item $\vec{Z}_{\mathrm{OA}_l^*}$ aligns with $\vec{Z}_\mathrm{S^*}$.
       \item $\vec{Y}_{\mathrm{OA}_l^*}$ is determined by the right-hand rule.
   \end{itemize}
\end{itemize}

It should be noted that the above definitions are purely geometric. For the purpose of our evaluation, we consider the effect of the finite light speed negligible, given the relatively short armlength of TianQin ($\sim$ 0.57 s light travel time). 

\subsection{Orbit-attitude coupling}\label{sec:OAdependence}
In order to calculate the nominal control states, analytical expressions of the angular velocities and accelerations of the target reference frames are derived, when the attitudes of the satellites are locked onto the constellation \cite{lupi2019precise, heisenberg2023lisa}. In order to simplify the expressions, we omit the subscript of $i$ when considering the multi-body dynamics of a single satellite. 

For basic notations, the satellite $i$'s position in the inertial frame is denoted by ${\vec{R}_i}$. The time derivatives of a unit vector $\vec{A}$ follow the identities
\begin{eqnarray}
    \dot{\vec{A}}  &=& {{\vec{\omega}}\times\vec{A}}, \\
    \ddot{\vec{A}} &=& {\dot{\vec{\omega}}}\times\vec{A}+{\vec{\omega}} \times ({\vec{\omega}}\times\vec{A}).
\end{eqnarray}
where $\vec{\omega}$ is the angular velocity.

The position of the constellation's incenter can be represented as 
\begin{equation}
    \vec{r}_{\mathrm{inc}} = \frac{{{r}_{23}}{\vec{R}_1}+{{r}_{31}}{\vec{R}_2}+{{r}_{12}}{\vec{R}_3}}{r_{12}+r_{23}+r_{31}}, 
\end{equation}
with
\begin{equation}
    \vec{r}_{ij}={\vec{R}_j}-{\vec{R}_i},\quad r_{ij}=|\vec{r}_{ij}|.
\end{equation}
Additionally, the breathing angle can be determined by 
\begin{equation}
    \alpha_i=\arccos(\vec n_{ij} \cdot \vec n_{ik}),
\end{equation}
with 
\begin{equation}
    \vec{n}_{ij}=\frac{\vec{r}_{ij}}{|\vec{r}_{ij}|}. 
\end{equation}

Now we can calculate the nominal attitudes of the satellites. First, $\vec{X}_{S_i^*}$ can be defined by
\begin{equation}
    \vec X_{\mathrm{S}_i^*} \equiv \frac{\vec{r}_{\mathrm{inc}}-\vec{R}_{i}}{|\vec{r}_{\mathrm{inc}}-\vec{R}_{i}|}, 
\end{equation}
and $\vec Z_{S_i^*}$ are given by
\begin{equation}
    \vec Z_{\mathrm{S}_1^*} = \vec{Z}_{\mathrm{S}_2^*} = \vec{Z}_{\mathrm{S}_3^*} \equiv \frac{\vec n_{ij} \times \vec n_{ik}}{\sin\alpha_i},
\end{equation}
where $[i, j, k]$ must be an even permutation of $[1, 2, 3]$. 

The coordinate transformation matrix from the $\mathcal{I}$ frame to ${\mathcal{S}_i^*}$ frame, for example, can be determined by
\begin{equation}
T_I^{S_i^*}=
\begin{bmatrix}
 \vec X_{S_i^*}^T \\
 \vec Y_{S_i^*}^T \\
 \vec Z_{S_i^*}^T
\end{bmatrix}, 
\end{equation}
where the direction column vectors of ${\mathcal{S}_i^*}$ are expressed in the $\mathcal{I}$ frame, and $\vec{A}^T$ means the transpose of the column vector $\vec{A}$. 

In order to obtain the angular velocity and acceleration of ${\mathcal{S}_i^*}$ with respect to $\mathcal{I}$, we introduce an auxiliary frame called the $p$-frame with the origin at the constellation's incenter $\vec{r}_{\mathrm{inc}}$ \cite{lupi2019precise}, and use the super- or subscript $p$ to annotate variables related to or expressed in the frame. The $z$-axis of the $p$-frame is
\begin{equation}
    \vec{z}_p = \vec{Z}_{S_i^*},
\end{equation}
and the $x$-axis aligns with the projection of $\vec{Z}_I$ onto the constellation plane, i.e.,
\begin{equation}
    \vec{x}_p=\frac{\cos{\beta}\vec{z}_p-\vec{Z}_I}{\sin\beta}, 
\end{equation}
with
\begin{equation}
    \cos\beta=\vec{Z}_I\cdot\vec{z}_p
\end{equation}
and $\vec{y}_p$ given by the right-hand rule. Hence, the angular velocity of the $p$-frame, expressed in its own coordinate system, is given by
\begin{equation}
\vec{\omega}_p=
\begin{bmatrix}
\vec{z}_p\cdot\dot{\vec{y}}_p\\
\vec{x}_p\cdot\dot{\vec{z}}_p\\
\vec{y}_p\cdot\dot{\vec{x}}_p
\end{bmatrix}.
\end{equation}
Now we can obtain the $\vec X_{S_i^*}$ vector of the satellite $i$ written in the $p$-frame as
\begin{equation}
    {^p\vec{x}_i}=-\frac{^p\vec{R}_i}{|^p\vec{R}_i|}
\end{equation}
with
\begin{equation}\label{eq:T_I^p}
    ^p\!\vec{R}_i=T_I^p({\vec{R}_i}-{\vec{r}_{\mathrm{inc}}}). 
\end{equation}
The angular velocity of the satellite $i$ with respective to the $p$ frame only have a nonzero $z$-component, so that one has
\begin{equation}
    ^p\vec{\omega}_{i/p}=\frac{^p\!\dot{\vec{R}}_i\times{^p\!\vec{x}_i}}{|^p\vec{R}_i|}.
\end{equation}
The angular acceleration of the $p$ frame can be written as 
\begin{equation}
    \dot{\vec{\omega}}_p=
    \begin{bmatrix}
    \dot{\vec{z}}_p\cdot\dot{\vec{y}}_p+\vec{z}_p\cdot\ddot{\vec{y}}_p\\
    \dot{\vec{x}}_p\cdot\dot{\vec{z}}_p+\vec{x}_p\cdot\ddot{\vec{z}}_p\\
    \dot{\vec{y}}_p\cdot\dot{\vec{x}}_p+\vec{y}_p\cdot\ddot{\vec{x}}_p
    \end{bmatrix},
\end{equation}
and the angular acceleration of the satellite $i$ with respective to the $p$ frame is given by
\begin{equation}
    ^p\dot{\vec{\omega}}_{i/p} = \frac{2({^p\vec{x}_i}\cdot{^p\!\dot{\vec{R}}_i}){^p\vec{\omega}_{i/p}}-{^p\vec{x}_i}\times{^p\!\ddot{\vec{R}}_i}}{|{^p\!\vec{R}_i}|}.
\end{equation}
Finally, the vectors ${^{S_i^*}\vec{\omega}_{i}}$ and ${^{S_i^*}\dot{\vec{\omega}}_{i}}$ can be represented as 
\begin{equation}\label{eq:T_p^{S_i^*}}
    {^{S_i^*}\vec{\omega}_{i}} = T_p^{S_i^*}({\vec{\omega}_{p}}+{^p\vec{\omega}_{i/p}})
\end{equation}
and
\begin{equation}
    {^{S_i^*}\dot{\vec{\omega}}_{i}} = T_p^{S_i^*}({\dot{\vec{\omega}}_{p}}+{^p\dot{\vec{\omega}}_{i/p}})+\dot{T}_p^{S_i^*}({\vec{\omega}_{p}}+{^p\vec{\omega}_{i/p}}).
\end{equation}

To summarize, a mathematical relation has been derived to provide the target attitudes, angular velocities and angular accelerations of the satellites, which are all determined from the orbit information of the three satellites. Likewise, similar relations can be derived for the MOSA's target frame $\mathcal{OA}_l^*$ by rotating ${\mathcal{S}_i^*}$ about the $\vec Z_{\mathrm{S}_i^*}$-axis. For readers' convenience, the Table \ref{tab:transMatrix} summarizes all the coordinate transformation matrices needed in this paper. 

\begin{table}[htb]
\renewcommand{\arraystretch}{1.5}
\caption{\label{tab:transMatrix} The summary of coordinate transformation matrices}
\begin{ruledtabular}
\begin{tabular}{c|c}
Symbols & Description \\ 
\hline
$T_I^{S_i^*}$ $(i=1,2,3)$ & $\text{From the inertial frame}$ \\
\quad & $\text{to the satellite $i$ target frame }$ \\
\hline
$T_I^{p}$ & $\text{From the inertial frame}$ \\
(see Eq. (\ref{eq:T_I^p})) & $\text{to the $p$-frame}$ \\
\hline
$T_p^{S_i^*}$ $(i=1,2,3)$ & $\text{From the $p$-frame}$ \\
(see Eq. (\ref{eq:T_p^{S_i^*}})) & $\text{to the satellite $i$ target frame }$ \\
\hline
$T^{OA_l}_{S}$ $(l=1,2)$ & $\text{From one satellite frame}$ \\
(see Eq. (\ref{eq:EOM_TM})) & $\text{to the corresponding MOSA $l$ frame }$ \\
\end{tabular}
\end{ruledtabular}
\end{table}

\subsection{Satellite model and thruster layout}\label{sec:SC_model}
Another focus of this research is to assess control requirements on the micro-propulsion, which executes the DFAC commands. In the science mode, the micro-propulsion subsystem is responsible for compensating non-conservative forces (mostly solar radiation pressure, SRP) on the satellite, and enabling the satellite to continuously track the two TMs along the sensitive axes, while maintaining the nominal attitudes and pointing with high precision. 

\begin{figure}[ht]
  \centering
  \includegraphics[width=0.48\textwidth]{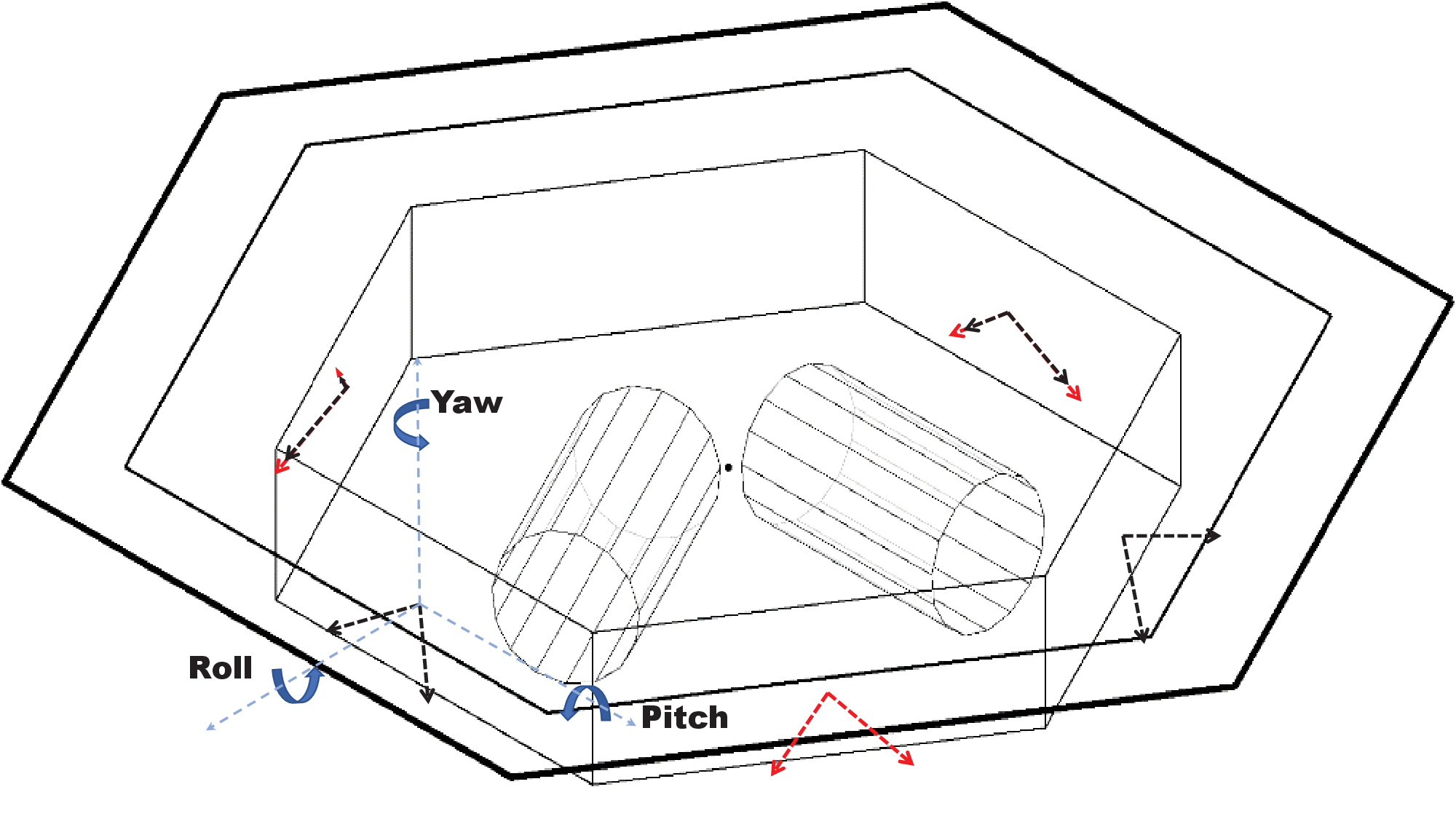}
\caption{\label{fig:ThrustersArrang} Illustration of the satellite model with two MOSAs (cylinders) inside. Thrusters are installed in a distributed and symmetric manner. Two options are to be evaluated, including four-cluster (black) and three-cluster (red) configurations, with each cluster having two diverging nozzles indicated by arrows.}
\end{figure}

The satellite body is modeled by a regular hexagonal prism with a side length of 1.5 m and a height of 0.6 m (see Fig. \ref{fig:ThrustersArrang}; for test and evaluation purposes only, not reflecting the final design \cite{zxf2018thermal, wang2023thermal}). The two cylinders inside represent the locations of the MOSAs. The flat-top sunshield is a hexagonal thin plate with a side length of 2.4 m, capable of preventing direct sunlight onto the satellite side panels at an incident angle of $45^\circ$. Some basic satellite parameters are given in the Table \ref{tab:SCparam}. 

For the micro-propulsion, the study considers two thruster configurations, i.e., a set of four clusters and a set of three clusters, and with each clusters containing two diverging nozzles (see Fig. \ref{fig:ThrustersArrang}). The installation must avoid obstructing the telescopes and plume impingement onto the satellite surfaces. The test layout is to put each cluster at the center of a side panel. The nozzles are all pointing away from the sunshield, and their directions can be adjusted to meet the control requirements and further optimized for fuel consumption. 

\begin{table}[htb]
\caption{\label{tab:SCparam} Satellite parameters used in the simulations}
\begin{ruledtabular}
\begin{tabular}{ccc}
Symbols & Parameters & Values \\ 
\hline
$m_\mathrm{S}$(kg) & $\text{Satellite mass}$ & $1000$ \\
$I_\mathrm{S}$(kg$\cdot$m$^2$) & $\text{Moment of inertia}$ & diag(583, 583, 1125) \\
\hline
$m_\mathrm{OA}$(kg) & $\text{MOSA mass}$ & $60$ \\
$I_\mathrm{OA}$(kg$\cdot$m$^2$) & $\text{Moment of inertia}$ & diag(1.52, 1.56, 1.56) \\
\hline
$m_\mathrm{TM}$(kg) & $\text{TM mass}$ & \hspace{1.4em}$2.45$ \cite{luo2016TianQin}\\
$I_\mathrm{TM}$(kg$\cdot$m$^2$) & $\text{Moment of inertia}$ & diag(0.001, 0.001, 0.001) \\
\hline
$\rho_a$ & $\text{Sunshield absorptivity}$ & $0.4$ \\
$\rho_r$ & $\text{Sunshield reflectivity}$ & \hspace{1.8em}$0.6$ \cite{wang2023thermal} \\
\end{tabular}
\end{ruledtabular}
\end{table}

\section{Nominal suspension Control of Test Masses}\label{sec:SuspensionC}
In \cite{vidano2020DYN}, the full equations of motion (EoM) of the TMs and satellites for LISA have been derived. It indicates the presence of differential inertial (e.g., centrifugal, Coriolis) accelerations introduced by the rotational motion of the satellite and MOSAs, as well as differential gravitational accelerations of the TMs and the satellite. These differential accelerations lead to relative motion among the TMs and the satellite, which must be compensated by the suspension control on TMs along the non-sensitive axes. 

Space-based GW detection requires that the differential self-gravity acceleration of the two TMs should be kept below $\sim 10^{-10}$ m/s$^2$ and the angular acceleration should be no more than $\sim 10^{-10}$ rad/s$^{-2}$ \cite{merkowitz2005self, armano2016constraints} to avoid excessive cross-talk of actuation noise to the sensitive axes ($\lesssim 10^{-15}$ m/s$^2$/Hz$^{1/2}$). Likewise, both differential inertial accelerations and differential gravitational accelerations should be also below this level, and their directions and magnitudes are affected by the positions of TMs and MOSA pivots relative to the CoM of the satellite, in addition to the satellite orbits. 

In this section, the nominal attitudes of the satellites and MOSAs will be incorporated with the EoM of the TMs to compute the required electrostatic control forces and torques for keeping the TMs aligned and at the centers of the EHs. Based on this framework, the placement of the TMs and MOSA pivots within the satellite can be optimized to lower the required control forces and torques. To focus on the orbit-related effects, we have excluded the self-gravity from the satellite in the studies. 

\subsection{Estimated electrostatic forces}\label{sec:Eforce}
In the science mode, electrostatic forces stabilize the two TMs along their non-sensitive axes relative to the EHs, while the DFAC and micro-propulsion of the satellite oversee the motion of the two TMs and sustain a stable dynamic relation with them \cite{amaro2017laser, ma2022controller}. Each TianQin satellite, situated in a geocentric orbit, experiences a stronger gravity gradient caused by the Earth-Moon system, when compared with LISA. Therefore, it is important to include this effect in calculating the required nominal control forces and torques for the satellites. The following calculation is based on rigid-body dynamics. 

The equation of $\mathrm{TM}_l$ $(l=1,2)$ translational motion is given by
\begin{equation}\label{eq:TM_l}
{^{I}\ddot{\vec{r}}_{\mathrm{TM}_{l}}} = {^{I}\vec{a}_{\mathrm{TM}_{l}}} + \frac{{^{I}\!{\vec{f}}_{\mathrm{c},l}}}{m_{\mathrm{TM}_{l}}} + \frac{{^{I}\!{\vec{f}}_{\mathrm{dis},l}}}{m_{\mathrm{TM}_{l}}},
\end{equation}
where ${^{I}\ddot{\vec{r}}_{\mathrm{TM}_{l}}}$ is the acceleration of $\mathrm{TM}_l$ with respective to the $\mathcal{I}$ frame, and the term $^{I}\vec{a}_{\mathrm{TM}_{l}}$ represents the gravitational acceleration of $\mathrm{TM}_l$, and $^{I}{\vec{f}_{\mathrm{c},l}}$ and ${^{I}{\vec{f}}_{\mathrm{dis},l}}$ are the control force and the disturbing force, respectively. 

To model the system accurately, one must formulate the system's dynamics in the reference frame where measurements are made. For example, the TM dynamics needs to be expressed in the frame attached to its own EH. Thereby, the full EoM of $\mathrm{TM}_l$ $(l=1,2)$ in the $\mathcal{OA}_l$ frame can be given by \cite{vidano2020DYN, lupi2019precise}
\begin{equation} \label{eq:EOM_TM}
\begin{split}
{^{OA_l}\ddot{\vec{r}}_{l}}& = +\frac{{^{OA_l}\vec{f}_{\mathrm{c},l}}}{m_{\mathrm{TM}_{l}}}+\frac{{^{OA_l}{\vec{f}}_{\mathrm{dis},l}}}{m_{\mathrm{TM}_{l}}}-\frac{{^{OA_l}\vec{f}_\mathrm{c,S}}}{m_\mathrm{S}}-\frac{{^{OA_l}\vec{f}_\mathrm{dis,S}}}{m_\mathrm{S}}
\\
&\quad +T^{OA_l}_{S}\{({^{S}\vec{a}_{\mathrm{TM}_{l}}}-{^{S}\vec{a}_\mathrm{S}})
\\
&\quad -{^{S}\ddot{\vec{r}}_{0,l}} - {^{S}\vec{\omega}_\mathrm{S}}\times[{^{S}\vec{\omega}_\mathrm{S}}\times({^{S}\vec{r}_{l}}+{^{S}{\vec{r}}_{0,l}})]
\\
&\quad -2 \; {^{S}\vec{\omega}_{S}}\times T^S_{OA_l}({^{OA_l}\dot{\vec{r}}_{l}}+{^{OA_l}\vec{\omega}_{\mathrm{OA}_l}}\times{^{OA_l}\vec{r}_{l}})
\\
&\quad -2 \; {^{S}\vec{\omega}_\mathrm{S}}\times{^{S}\dot{\vec{r}}_{0,l}} - {^{S}\dot{\vec{\omega}}_\mathrm{S}}\times({^{S}\vec{r}_{l}}+{^{S}{\vec{r}}_{0,l}})\}\\
&\quad -{^{OA_l}\vec{\omega}_{\mathrm{OA}_l}}\times({^{OA_l}\vec{\omega}_{\mathrm{OA}_l}}\times{^{OA_l}\vec{r}_{l}})\\
&\quad -2 \; {^{OA_l}\vec{\omega}_{\mathrm{OA}_l}}\times{^{OA_l}\dot{\vec{r}}_{l}}-{^{OA_l}\dot{\vec{\omega}}_{\mathrm{OA}_l}}\times{^{OA_l}\vec{r}_{l}}.
\end{split}
\end{equation}
Here ${^{OA_l}\ddot{\vec{r}}_{l}}$ is the acceleration of $\mathrm{TM}_l$ with respective to the $\mathcal{OA}_l$ frame. The terms ${^{S}\vec{a}_{\mathrm{TM}_{l}}}$ and ${^{S}\vec{a}_{S}}$ are the gravitational acceleration of $\mathrm{TM}_l$ and the satellite in the $\mathcal{S}$ frame, respectively. ${^{OA_l}\vec{f}_\mathrm{c,S}}$ and ${^{OA_l}\vec{f}_\mathrm{dis,S}}$ are the control forces and the disturbing force on the satellite in the $\mathcal{OA}_l$ frame. The symbol $T^{OA_l}_{S}$ is the transformation matrix from the $\mathcal{S}$ frame to the $\mathcal{OA}_l$ frame. The terms ${^{S}\vec{\omega}_\mathrm{S}}$ and ${^{S}\dot{\vec{\omega}}_\mathrm{S}}$ denote the angular velocity and acceleration of the satellite, respectively, and likewise, the terms ${^{S}\vec{\omega}_{\mathrm{OA}_l}}$ and ${^{S}\dot{\vec{\omega}}_{\mathrm{OA}_l}}$ denote MOSA's angular velocity and acceleration, respectively. The terms ${^{S}\dot{\vec{r}}_{0,l}}$ and ${^{S}\ddot{\vec{r}}_{0,l}}$ can be expressed as
\begin{equation} \label{eq:r0Ldot}
    {^{S}\dot{\vec{r}}_{0,l}} = {^{S}\vec{\omega}_{\mathrm{OA}_l}}\times( {^{S}{\vec{r}}_{0,l}}-{^{S}{\vec{r}}_{\mathrm{p},l}})
\end{equation}
and
\begin{equation} \label{eq:rOLddot}
\begin{aligned}
    {^{S}\ddot{\vec{r}}_{0,l}} &= {^{S}\dot{\vec{\omega}}_{\mathrm{OA}_l}}\times( {^{S}{\vec{r}}_{0,l}}-{^{S}{\vec{r}}_{\mathrm{p},l}})
    \\
    &\quad +{^{S}\vec{\omega}_{\mathrm{OA}_l}}\times[{^{S}\vec{\omega}_{\mathrm{OA}_l}}\times( {^{S}{\vec{r}}_{0,l}}-{^{S}{\vec{r}}_{\mathrm{p},l}})],
\end{aligned}
\end{equation}
where ${^{S}{\vec{r}}_{\mathrm{p},l}}$, $^{S}{\vec{r}}_{0,l}$ are the MOSA's pivot position and the EH's center position with respective to the satellite. 

Under nominal control, the disturbing forces are compensated by the control forces, and for simplicity we absorb ${^{OA_l}{\vec{f}}_{\mathrm{dis},l}}$ and ${^{OA_l}{\vec{f}}_{\mathrm{dis},\mathrm{S}}}$ into $^{OA_l}{\vec{f}}_{\mathrm{c},l}$ and $^{OA_l}{\vec{f}}_{\mathrm{c},\mathrm{S}}$, respectively. Moreover, there is no relative motion between the TM and the EH, i.e., $^{OA_l}\vec{r}_{l} = {^{OA_l}\dot{\vec{r}}_{l}} = {^{OA_l}\ddot{\vec{r}}_{l}} = 0$. Hence, from Eq. (\ref{eq:EOM_TM}) with Eqs. (\ref{eq:r0Ldot}) and (\ref{eq:rOLddot}) plugged in, the equation determining the nominal suspension control force $^{OA_l}{\vec{f}}_{\mathrm{c},l}$ can be written as
\begin{equation}\label{eq:TMforce}
\begin{split}
\frac{^{OA_l}{\vec{f}}_{\mathrm{c},l}}{m_{\mathrm{TM}_{l}}} & = \frac{^{OA_l}{\vec{f}}_{\mathrm{c,S}}}{m_\mathrm{S}} - T^{OA_l}_{S}\{({^{S}\vec{a}_{\mathrm{TM}_{l}}}-{^{S}\vec{a}_{\mathrm{S}}})
\\
&\quad -2 \; {^{S}\vec{\omega}_\mathrm{S}}\times[{^{S}\vec{\omega}_{\mathrm{OA}_l}}\times({^{S}{\vec{r}}_{0,l}}-{^{S}{\vec{r}}_{\mathrm{p},l}})]
\\
&\quad -{{^{S}\dot{\vec{\omega}}_{\mathrm{OA}_l}}\times({^{S}{\vec{r}}_{0,l}}-{^{S}{\vec{r}}_{\mathrm{p},l}})}
\\
&\quad -{^{S}\vec{\omega}_{\mathrm{OA}_l}}\times[{^{S}\vec{\omega}_{\mathrm{OA}_l}}\times({^{S}{\vec{r}}_{0,l}}-{^{S}{\vec{r}}_{\mathrm{p},l}})] 
\\
&\quad -{{^{S}\vec{\omega}_\mathrm{S}}\times({^{S}\vec{\omega}_\mathrm{S}}\times{^{S}{\vec{r}}_{0,l}})} - {^{S}\dot{\vec{\omega}}_\mathrm{S}}\times{^{S}{\vec{r}}_{0,l}}\},
\end{split}
\end{equation}
and the above equation can be rewritten in a simpler form,  
\begin{equation}\label{eq:TMforcesimple}
 {T_S^{OA_l}} \: {^{S}\!{\vec{A}}_{\mathrm{c,S}}} = {^{OA_l}{\vec{A}}_{\mathrm{c},l}} + T^{OA_l}_{S} \: {^{S}\vec{g}_{l}}.
\end{equation}
Here ${^{OA_l}{\vec{A}}_{\mathrm{c},l}} := {^{OA_l}}\!{\vec{f}}_{\mathrm{c},l}/{m_{\mathrm{TM}_{l}}}$ is the suspension control acceleration on TM$_l$, and ${^{S}\!{\vec{A}}_{\mathrm{c,S}}} := {^{S}}\!{\vec{f}}_{\mathrm{c,S}}/{m_\mathrm{S}}$ is the control acceleration the satellite needs to follow the TMs. Moreover, ${^{S}\vec{g}_l}$ is the TM acceleration in the $\mathcal{S}$ frame given by the terms in the curly brackets of Eq. (\ref{eq:TMforce}). Now the key step is to equalize the accelerations of the two TMs by the suspension control \cite{inchauspe2022new}, as follows. 

First, the differential acceleration $\Delta^{S}\vec{g}$ of the two TMs is defined by
\begin{equation}\label{eq:E}
    {\Delta {^{S}\vec{g}}} = {^{S}\vec{g}_{2}}-{^{S}\vec{g}_{1}}.
\end{equation}
Moreover, we use $^S\vec{G}_l$ to denote the accelerations of the two TMs under the suspension control in the $\mathcal{S}$ frame, and we obtain 
\begin{equation}\label{eq:GE}
\begin{aligned}
    ^S\vec{G}_1 = {^{S}\!\vec{A}_{\mathrm{c},1}} + {^{S}\!\vec{g}_{1}},\\
    ^S\vec{G}_2 = {^{S}\!\vec{A}_{\mathrm{c},2}} + {^{S}\!\vec{g}_{2}}.
\end{aligned}
\end{equation}

Second, we introduce $\theta$ and $\varphi$ as the angles between the optical axes $\vec{X}_{\mathrm{OA}_l}$ and $\vec{X}_{S}$, respectively (see Fig. \ref{fig:SC}). There are three conditions to be satisfied by the nominal DFAC, i.e., a) both TMs having equal accelerations in the $\mathcal{S}$ frame, i.e., $^S\vec{G}_1$=$^S\vec{G}_2$, and b) no suspension implemented along the sensitive axes $\vec{X}_{\mathrm{OA}_l}$, and c) the control forces along the $z$-axes of the two TMs being equal with opposite signs. The last condition is needed to compensate for the differential accelerations of the two TMs. Therefore, we can obtain the required electrostatic acceleration on $\mathrm{TM}_l$ in the $\mathcal{OA}_l$ frame as
\begin{equation}\label{eq:E1}
   {^{OA_1}{\vec{A}}_{\mathrm{c},1}}=\frac{\Delta{^{S}\!g_\mathrm{y}}\sin{\varphi}-\Delta{^{S}\!g_\mathrm{x}}\cos{\varphi}}{\sin(\varphi+\theta)}{\vec{Y}_{\mathrm{OA}_1}} + \frac{\Delta{^{S}\!g_\mathrm{z}}}{2}{\vec{Z}_{\mathrm{OA}_1}},
\end{equation}
\begin{equation}\label{eq:E2}
    {^{OA_2}{\vec{A}}_{\mathrm{c},2}}=\frac{-\Delta{^{S}\!g_\mathrm{y}}\sin{\theta} - \Delta{^{S}\!g_\mathrm{x}}\cos{\theta}}{\sin(\theta+\varphi)}{\vec{Y}_{\mathrm{OA}_2}} - \frac{\Delta {^{S}\!g_\mathrm{z}}}{2}{\vec{Z}_{\mathrm{OA}_2}},
\end{equation}
where $\Delta{^{S}\!g_\mathrm{x}}$, $\Delta{^{S}\!g_\mathrm{y}}$, and $\Delta{^{S}\!g_\mathrm{z}}$ are the components of ${\Delta{^{S}\!\vec{g}}}$ in the $\mathcal{S}$ frame. 

Finally, taking Eq. (\ref{eq:E1}) or Eq. (\ref{eq:E2}) back to Eq. (\ref{eq:GE}), the common acceleration of the two TMs under the suspension control in the $\mathcal{S}$ frame reads
\begin{equation}\label{eq:G}
\begin{aligned}
    ^S\vec{G} & = \frac{{\sin\theta}{\cos\varphi} \; {^{S}\!{g}_{\mathrm{x},2}} + {\cos\theta}{\sin\varphi} \; {^{S}\!{g}_{\mathrm{x},1}}-{\sin\theta}{\sin\varphi} \; {\Delta{^{S}\!g_\mathrm{y}}}}{\sin(\theta+\varphi)}\vec{X}_\mathrm{S}
    \\
    & +\frac{{\sin\varphi}{\cos\theta} \; {^{S}\!{g}_{\mathrm{y},2}}+{\cos\varphi}{\sin\theta} \; {^{S}\!{g}_{\mathrm{y},1}}-{\cos\theta}{\cos\varphi} \; {\Delta{^{S}\!g_\mathrm{x}}}}{\sin(\theta+\varphi)}\vec{Y}_\mathrm{S}
    \\
    & +\frac{g_{\mathrm{z},2}+g_{\mathrm{z},1}}{2}\vec{Z}_\mathrm{S},
\end{aligned}
\end{equation}
which is also the control acceleration of the satellite, i.e., ${^{S}\!{\vec{A}}_{\mathrm{c,S}}} = {^S\vec{G}}$. 

\begin{figure}[ht]
\centering
\includegraphics[width=0.48\textwidth]{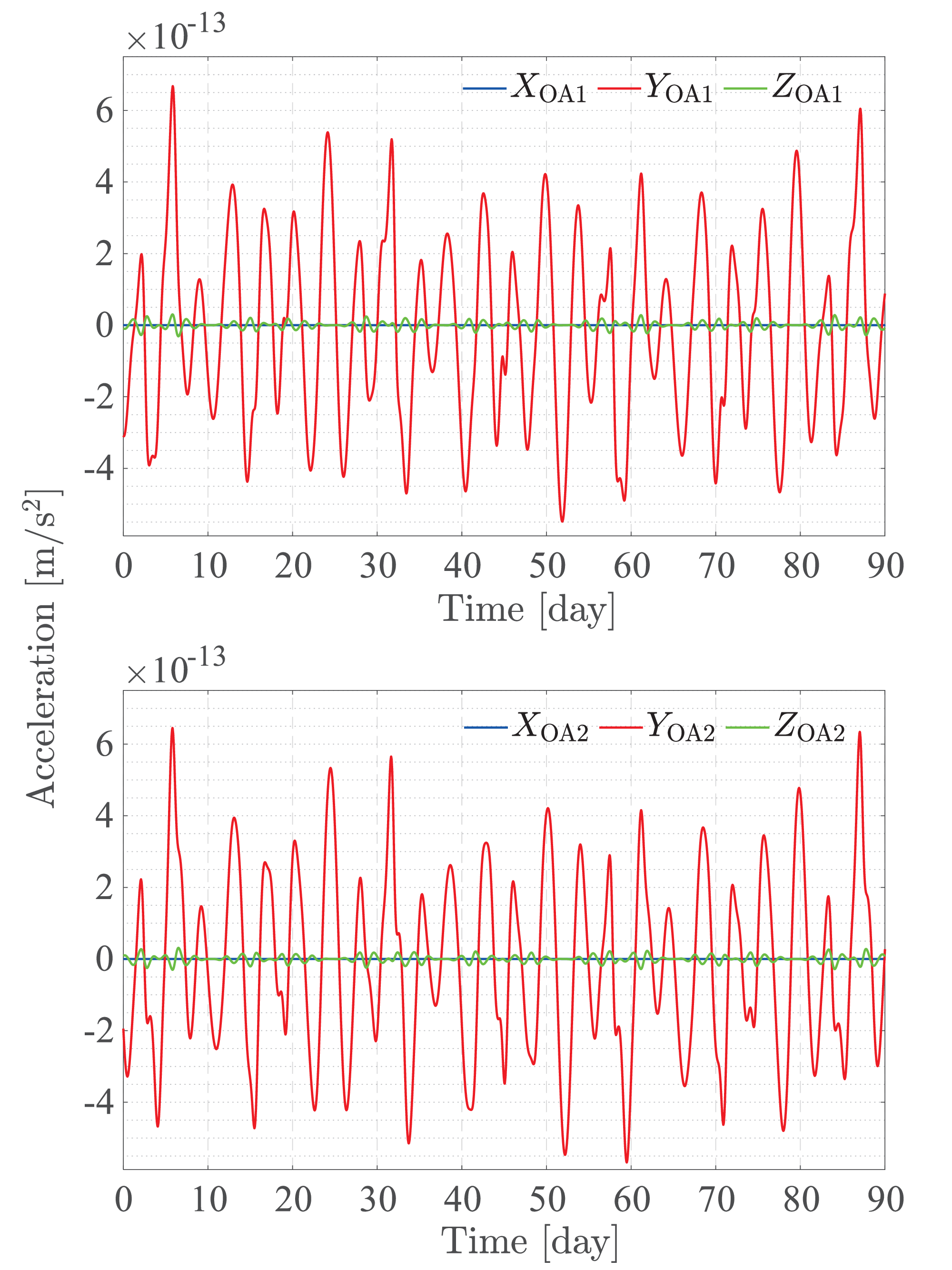}
\caption{\label{fig:DFforce}The nominal electrostatic control acceleration required for the two TMs in the science mode. }
\end{figure}

In the nominal control, the target frames and the body-fixed frames are aligned. So in the subsequent discussions, we will no longer differentiate between them. Now one can compute the electrostatic control accelerations for the TMs in the science mode, with the information of gravitational forces and the satellite/MOSA attitudes fed into the expressions of ${^{S}\vec{g}_l}$. Given the EH centers at $(0,20,0)$ cm and $(0,-20,0)$ cm in the $\mathcal{S}$ frame, the result on TM$_1$ is shown in Fig. \ref{fig:DFforce}. One can see that the control acceleration in the $\mathcal{OA}_1$ frame is zero along the $x$-axis, and in the order of $10^{-13}$ m/s$^2$ along the $y$-axis, and in the order of $10^{-14}$ m/s$^2$ along the $z$-axis. These values are well below the $10^{-10}$ m/s$^2$ requirement. 

The above calculations are under the condition of compensating the breathing angle through symmetric rotation of the MOSAs, i.e., $\theta=\phi$. We have also calculated the asymmetric case with one MOSA fixed, and it gives a similar result with no changes in the order of magnitude and hence omitted here. 

\begin{figure}[ht]
\centering
\includegraphics[width=0.48\textwidth]{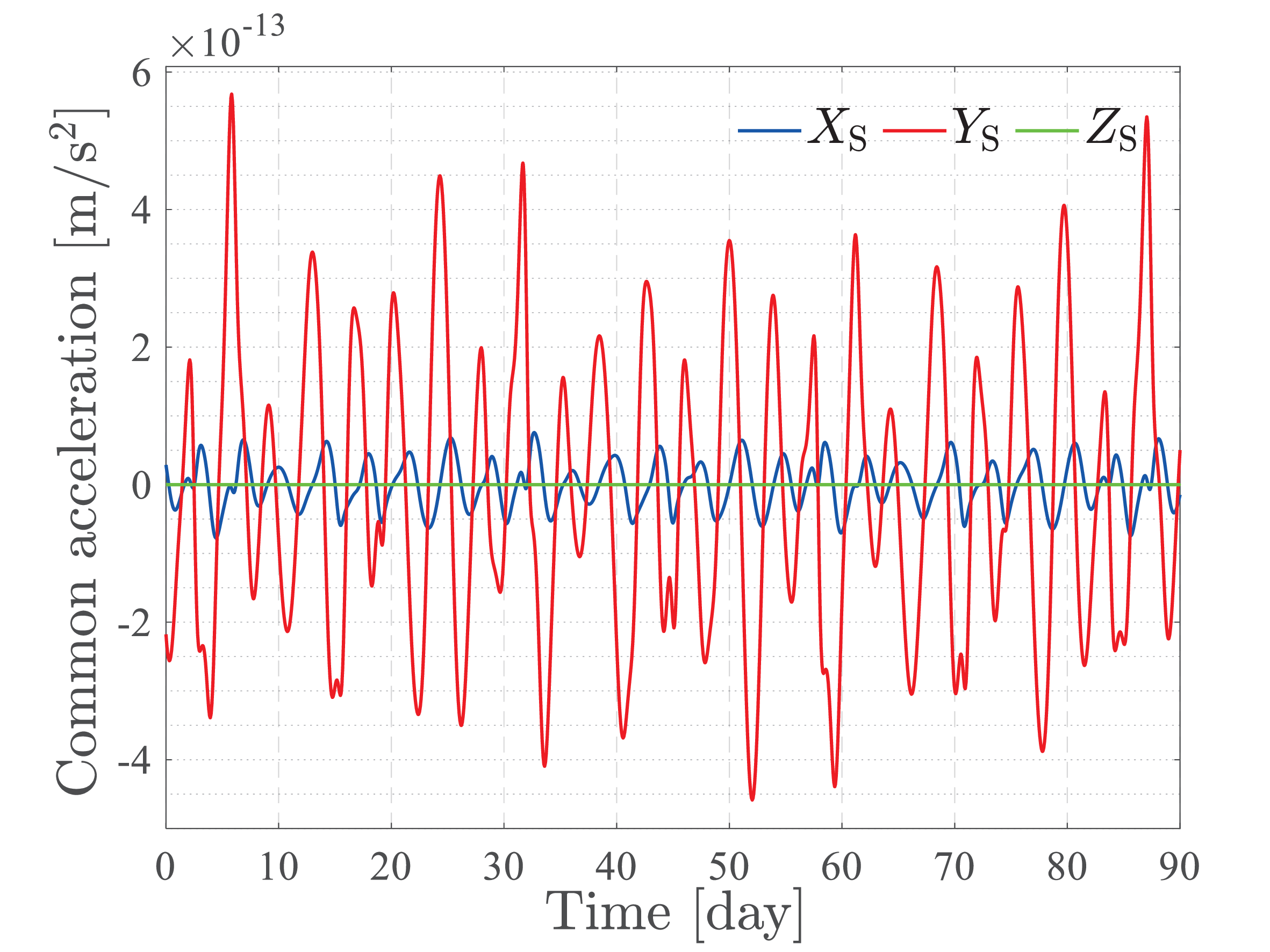}
\caption{\label{fig:commonG}The control acceleration required on the satellite, which is to be achieved by micro-propulsion. }
\end{figure}

Finally, we point out that the common acceleration $^SG$ is in the order of $10^{-13}$ m/s$^2$ (see Fig. \ref{fig:commonG}). Note that the control acceleration has zero values along the satellite's $z$-axis due to the condition c) mentioned earlier. In the Appendix \ref{App:orbit_dev}, orbital calculations with this acceleration added reveal negligible deviations from pure gravity orbits over three months, thus validating the assumption made in Sec. \ref{sec:orbit}. 

\subsection{Estimated electrostatic torques}\label{sec:Etorque}

\begin{figure}[ht]
\centering
\includegraphics[width=0.48\textwidth]{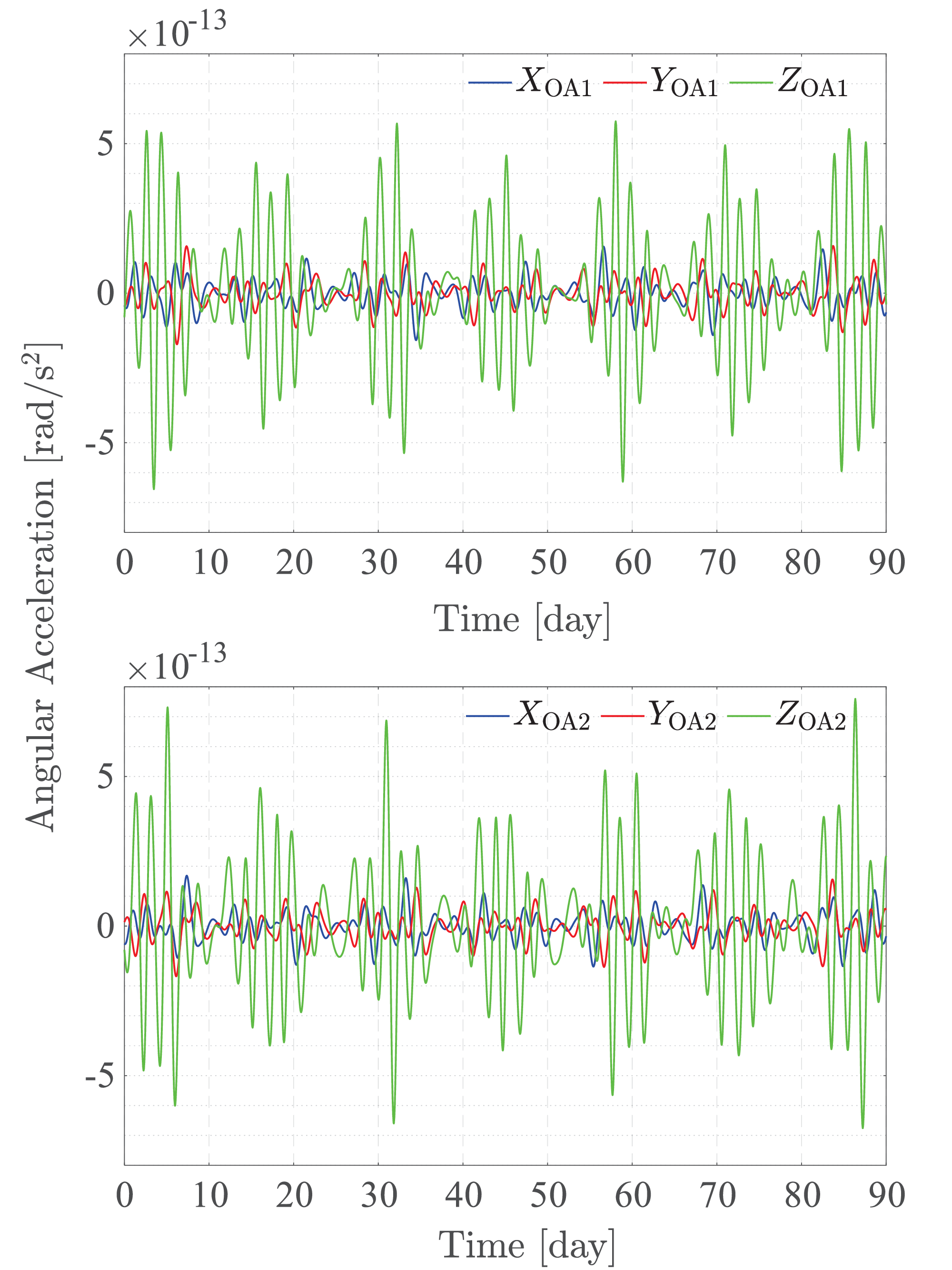}
\caption{\label{fig:Etorque} The nominal electrostatic angular accelerations of the two TMs in the science mode for three months. }
\end{figure}

For the inter-satellite measurements, the TMs need to be rotated with the satellite and MOSAs. Therefore, electrostatic control torques are applied to maintain the nominal attitudes of the TMs. The Euler's rotation equations for the TMs are given by
\begin{equation}
\begin{aligned}
{^{OA_l}\dot{\vec{\omega}}_{\mathrm{TM}_{l}}} &= {I_{\mathrm{TM}_l}^{-1}}({^{OA_l}{\vec{M}}_{\mathrm{c},l}} + {^{OA_l}{\vec{M}}_{\mathrm{dis},l}})
\\
&\quad -{I_{\mathrm{TM}_l}^{-1}}[{^{OA_l}{\vec{\omega}}_{\mathrm{TM}_{l}}} \times (I_{\mathrm{TM}_l} \: {^{OA_l} {\vec{\omega}}_{\mathrm{TM}_{l}}})].
\end{aligned}
\end{equation}
The terms ${^{OA_l}{\vec{M}}_{\mathrm{c},l}}$ and ${^{OA_l}{\vec{M}}_{\mathrm{dis},l}}$ are the control torques and the disturbance torques. The terms ${^{OA_l}{\vec{\omega}}_{\mathrm{TM}_{l}}}$ and  ${^{OA_l}\dot{\vec{\omega}}_{\mathrm{TM}_{l}}}$ are the angular velocity and acceleration of TM$_l$ in the $\mathcal{OA}_l$ frame with respective to the $\mathcal{I}$ frame, which can be written as
\begin{equation}
\begin{aligned}
{^{OA_l}{\vec{\omega}}_{\mathrm{TM}_{l}}}& = T^{OA_l}_{S} \: {^{S}\!\vec{\omega}_\mathrm{S}}+{^{OA_l}\vec{\omega}_{\mathrm{OA}_l}}+{^{OA_l}\vec{\omega}_{l}},
\end{aligned}
\end{equation}
and
\begin{equation}
\begin{aligned}
{^{OA_l}\dot{\vec{\omega}}_{\mathrm{TM}_{l}}}& = T^{OA_l}_{S}\:{^{S}\!\dot{\vec{\omega}}_\mathrm{S}}-{^{OA_l}\vec{\omega}_{\mathrm{OA}_l}} \times (T^{OA_l}_{S}\:{^{S}\!\vec{\omega}_\mathrm{S}})
\\
&\quad +{^{OA_l}\dot{\vec{\omega}}_{\mathrm{OA}_l}}+{^{OA_l}\dot{\vec{\omega}}_{l}}.
\end{aligned}
\end{equation}
Similar to the case of the nominal control accelerations, the relative angular velocities ${^{OA_l}{\vec{\omega}}_{l}}$ and accelerations ${^{OA_l}\dot{\vec{\omega}}_{l}}$ between TMs and their EHs, and the disturbance torques should all be zero. Hence the equations for the nominal electrostatic control torques read
\begin{equation}\label{eq:TMtorque}
\begin{aligned}
{^{OA_l}{\vec{M}}_{\mathrm{c},l}}& = I_{\mathrm{TM}_l} ({^{OA_l}\dot{\vec{\omega}}_\mathrm{S}} - {^{OA_l}\vec{\omega}_{\mathrm{OA}_l}} \times {^{OA_l}\vec{\omega}_{S}} + {^{OA_l}\dot{\vec{\omega}}_{\mathrm{OA}_l}})
\\
&\quad +({^{OA_l}\vec{\omega}_\mathrm{S}} + {^{OA_l}\vec{\omega}_{\mathrm{OA}_l}}) \times I_{\mathrm{TM}_l}\:{^{OA_l}\vec{\omega}_\mathrm{S}}
\\
&\quad +({^{OA_l}\vec{\omega}_\mathrm{S}}+{^{OA_l}\vec{\omega}_{\mathrm{OA}_l}}) \times I_{\mathrm{TM}_l}\:{^{OA_l}\vec{\omega}_{\mathrm{OA}_l}}.
\end{aligned}
\end{equation}
The above equations have no dependence on the positions of the EH centers and MOSA pivots, since the nominal attitudes are only determined by the orbits of the CoMs of the satellites in our treatment (see Sec. \ref{sec:OAdependence}). We can convert the torques to the angular accelerations, which are more convenient for comparing with the requirements. The electrostatic angular accelerations on the TMs over three months are shown in Fig. \ref{fig:Etorque}. It can be seen that the maximum angular acceleration for the two TMs is less than $10^{-12}$ rad/s$^2$, which are well below the requirement. 

\subsection{Optimizing TM Placement}\label{sec:TM_arrangement}
In the two previous subsections, calculations have been performed to determine the nominal electrostatic forces and torques on the two TMs. This subsection aims to find an optimal TM layout that minimizes the nominal control forces, providing a reference for the key payload and satellite design. 

Since the two TMs are symmetrically positioned relative to the $\vec{X}_\mathrm{S}$-$\vec{Z}_\mathrm{S}$ plane and are subject to similar dynamical environment, we focus on TM$_1$ and show its results in the following analysis. $\mathrm{TM}_1$ can be placed within the range $Y_\mathrm{S}\in\left[0,50\right]$ cm, $X_\mathrm{S}\in\left[-50,50\right]$ cm, and $Z_\mathrm{S}\in\left[-20,20\right]$ cm in the $\mathcal{S}$ frame. The TM positions are sampled at 5 cm intervals along $Y_\mathrm{S}$ and $X_\mathrm{S}$, and sampled at 10 cm intervals along $Z_\mathrm{S}$. The control accelerations at these positions are calculated and compared to identify the optimal position.  

\begin{figure}[ht]
\begin{minipage}{0.48\textwidth}
  \centering
  \includegraphics[width=\textwidth]{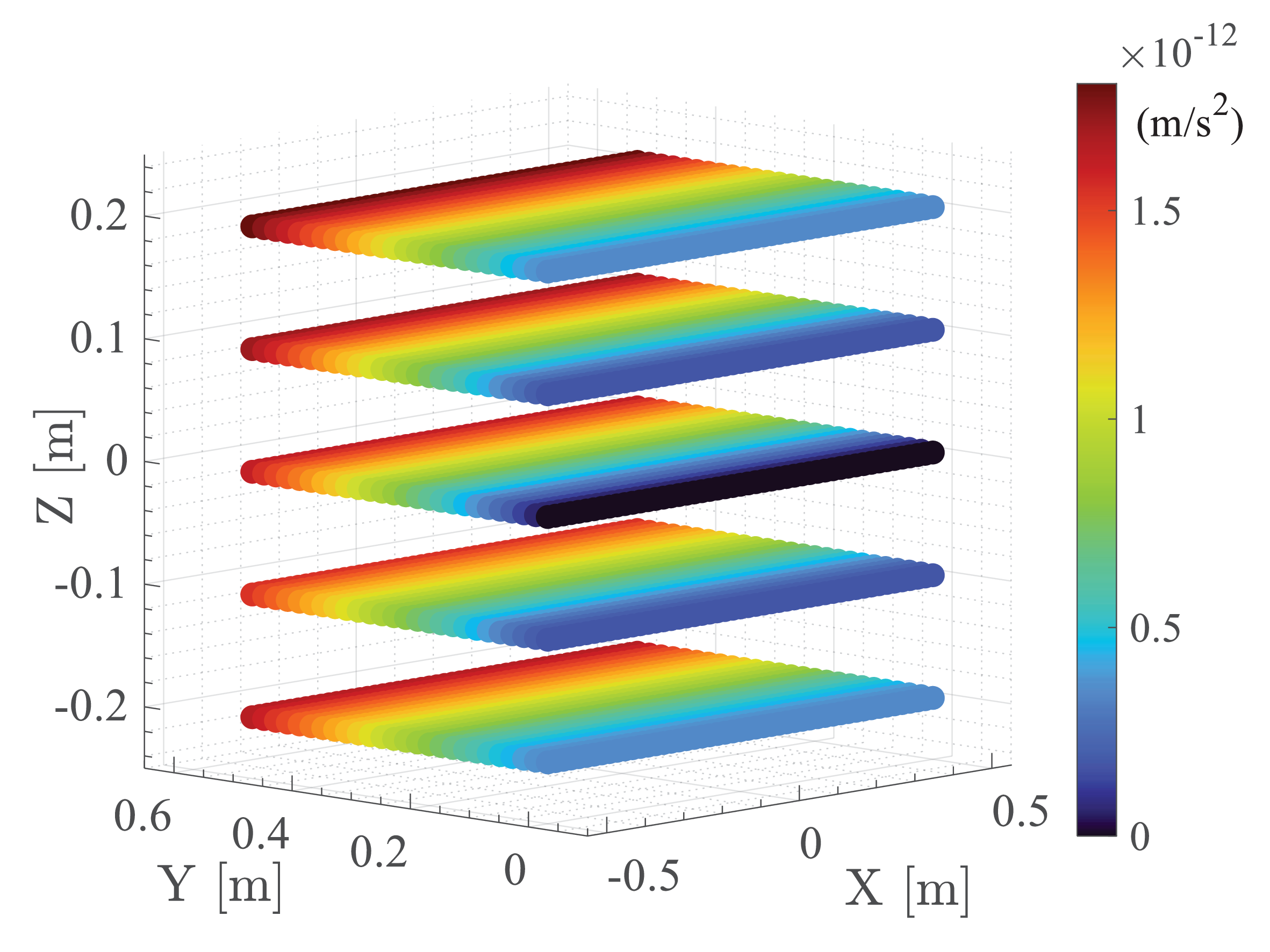}
\end{minipage}
\begin{minipage}{0.48\textwidth}
  \centering
  \includegraphics[width=\textwidth]{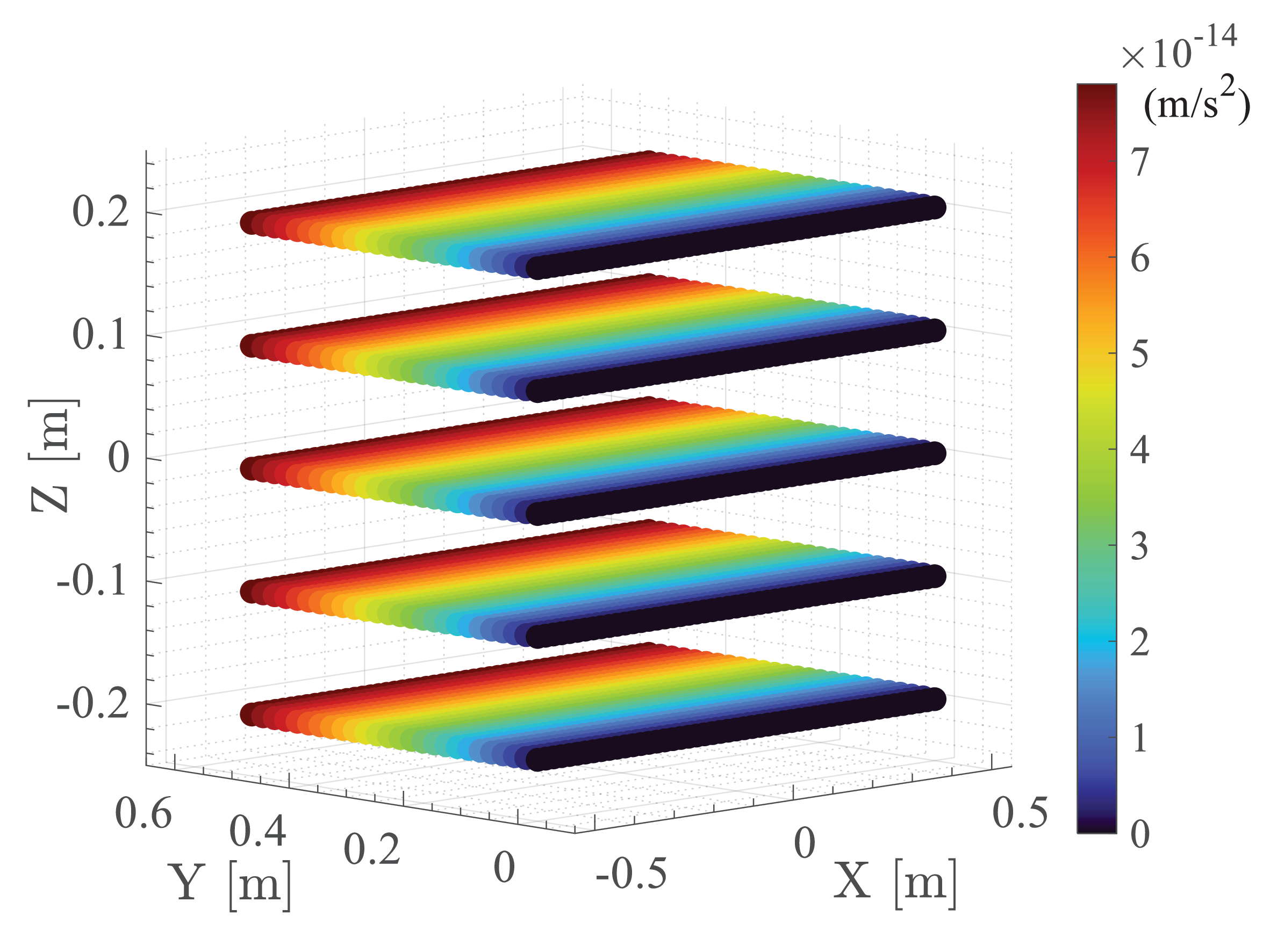}
\end{minipage}
\caption{\label{fig:TMArrangDF} The dependence of the maximum nominal TM control accelerations during three months on the TM$_1$ position. The upper plot shows the $\vec{Y}_{\mathrm{OA}_1}$ component, and the lower the $\vec{Z}_{\mathrm{OA}_1}$ component. }
\end{figure}

The dependence of the maximum (absolute) nominal TM control accelerations during three months on the TM$_1$ position are demonstrated in Fig. \ref{fig:TMArrangDF}. The plots show that the control accelerations remain nearly invariant with the TM positions shifting along the $\vec{X}_\mathrm{S}$. This is related to the fact that the sensitive axes of the TMs are not actuated. Furthermore, as the TM separation increases, i.e., shifting along $\vec{Y}_\mathrm{S}$, the differential gravitational acceleration grows, thereby augmenting the electrostatic control acceleration. However, it is still well below the requirement within a separation up to 1 m. In the case of placement along $\vec{Z}_\mathrm{S}$, the differential gravitational acceleration remains nearly constant, but the inertial acceleration varies slightly, resulting in a small elevation of the electrostatic acceleration when moving away from $\vec{Z}_\mathrm{S}=0$. To summarize, electrostatic control along the TM non-sensitive axes prefer shortening the TM separation, but generally it does not import strong limitation on the TM placement for TianQin, if one disregards the effect of self-gravity. 

\begin{figure}[ht]
    \centering
    \includegraphics[width=0.48\textwidth]{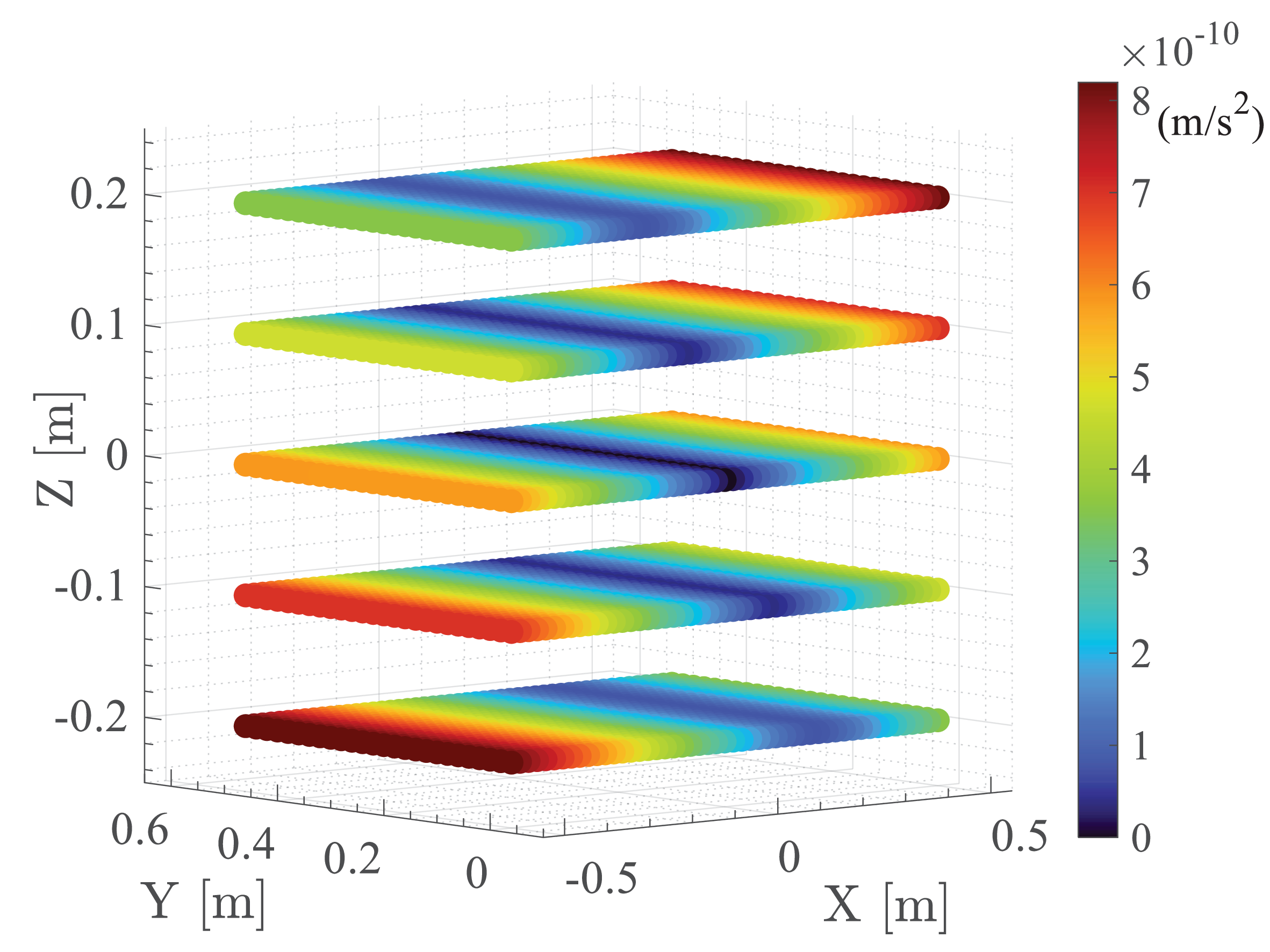}
    \caption{The dependence of the maximum nominal satellite control accelerations ($|^S\vec{G}|$) during three months on the TM$_1$ position. 
    }
    \label{fig:TMArrangG}
\end{figure}

Nevertheless, the TM placement along the $\vec{X}_\mathrm{S}$ axis can make a difference for the satellite control. As the Fig. \ref{fig:TMArrangG} shows, the required satellite acceleration ($|^S\vec{G}|$) to follow TMs is minimized when both CoMs of the TMs are placed in the $X_\mathrm{S}$ axis, i.e., being collinear and equidistant with the satellite CoM. This is important for saving fuel during the science observation. 

\subsection{Optimizing MOSA pivot placement} \label{sec:Optimalaxes}
\begin{figure}[ht]
  \centering
  \includegraphics[width=0.48\textwidth]{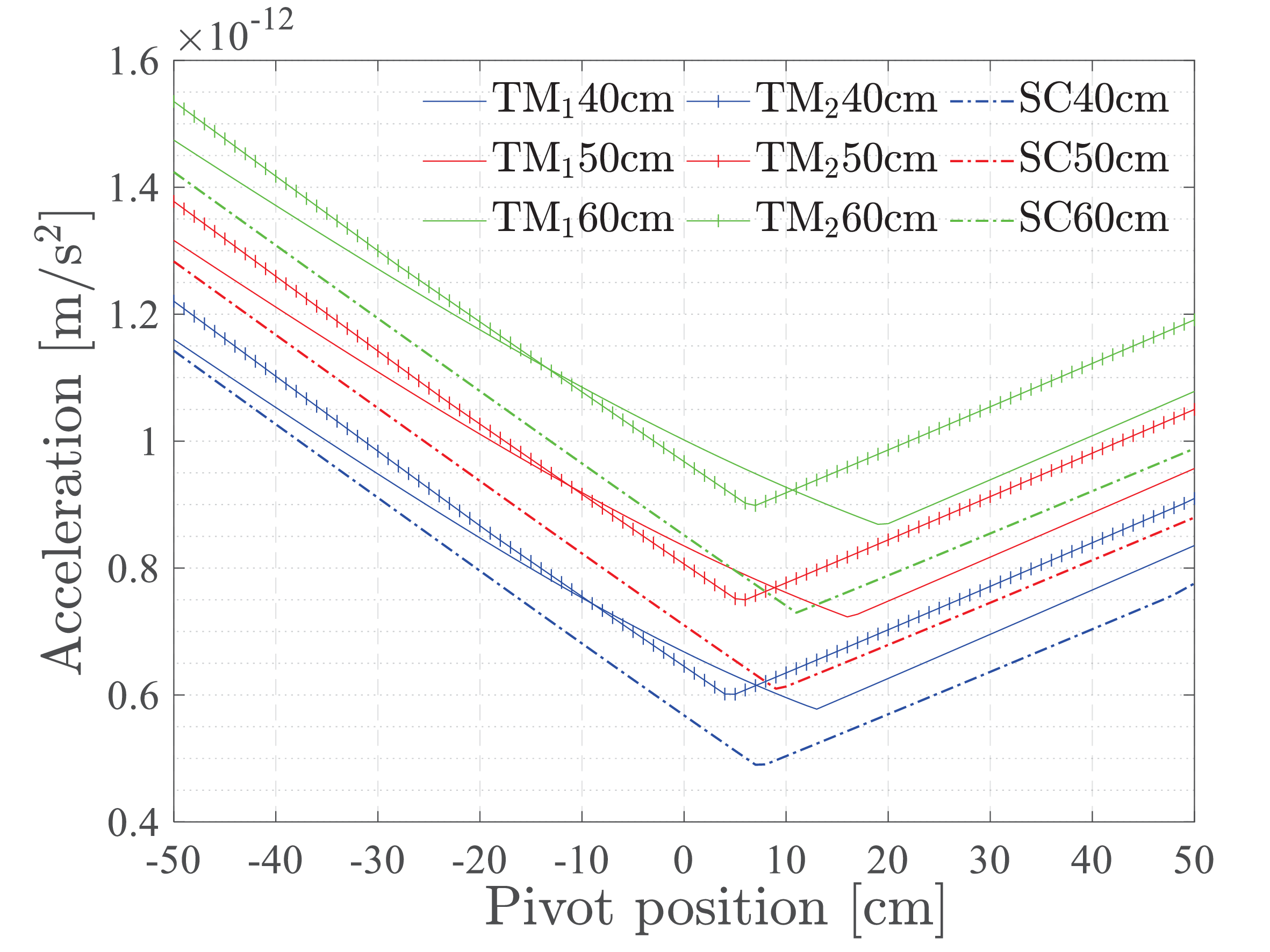}
\caption{\label{fig:axesPOS} The dependence of maximum nominal control accelerations of the TMs and satellite on the pivot positions, with different TM separations (40, 50, 60 cm). }
\end{figure}

We further examine the dependence of the maximum (absolute) nominal TM/satellite control accelerations during three months on the MOSA pivot positions. The pivot positions are sampled at 1 cm intervals within a range of $\left[-50,50\right]$ cm, relative to the EH center along the sensitive axes $\vec{X}_{\mathrm{OA}_l}$, and the TMs are separated by 40, 50, and 60 cm and aligned with the satellite CoM. Since the electrostatic control ${^{OA_l}{\vec{A}}_{\mathrm{c},l}}$ occurs predominantly along $\vec{Y}_{\mathrm{OA}_l}$, we only show the results in the these directions of the TMs, and for the satellite we show the magnitude of the control acceleration, all in Fig. \ref{fig:axesPOS}. From the plot, we note that $\mathrm{TM}_2$ reaches its minimum value at a pivot position different from $\mathrm{TM}_1$. This is owing to that the control accelerations of the two TMs are different as shown in Eq. (\ref{eq:E1}) and Eq. (\ref{eq:E2}). Therefore one can find a pivot position to minimize the averaged maximum control accelerations of the two TMs. Thus the position of approximately 10 cm ahead of the EH center is advisable. Moreover, the asymmetrical V-shaped general trend is similar for different TM separations and for the satellite.
 
\begin{figure}[ht]
  \centering
  \includegraphics[width=0.48\textwidth]{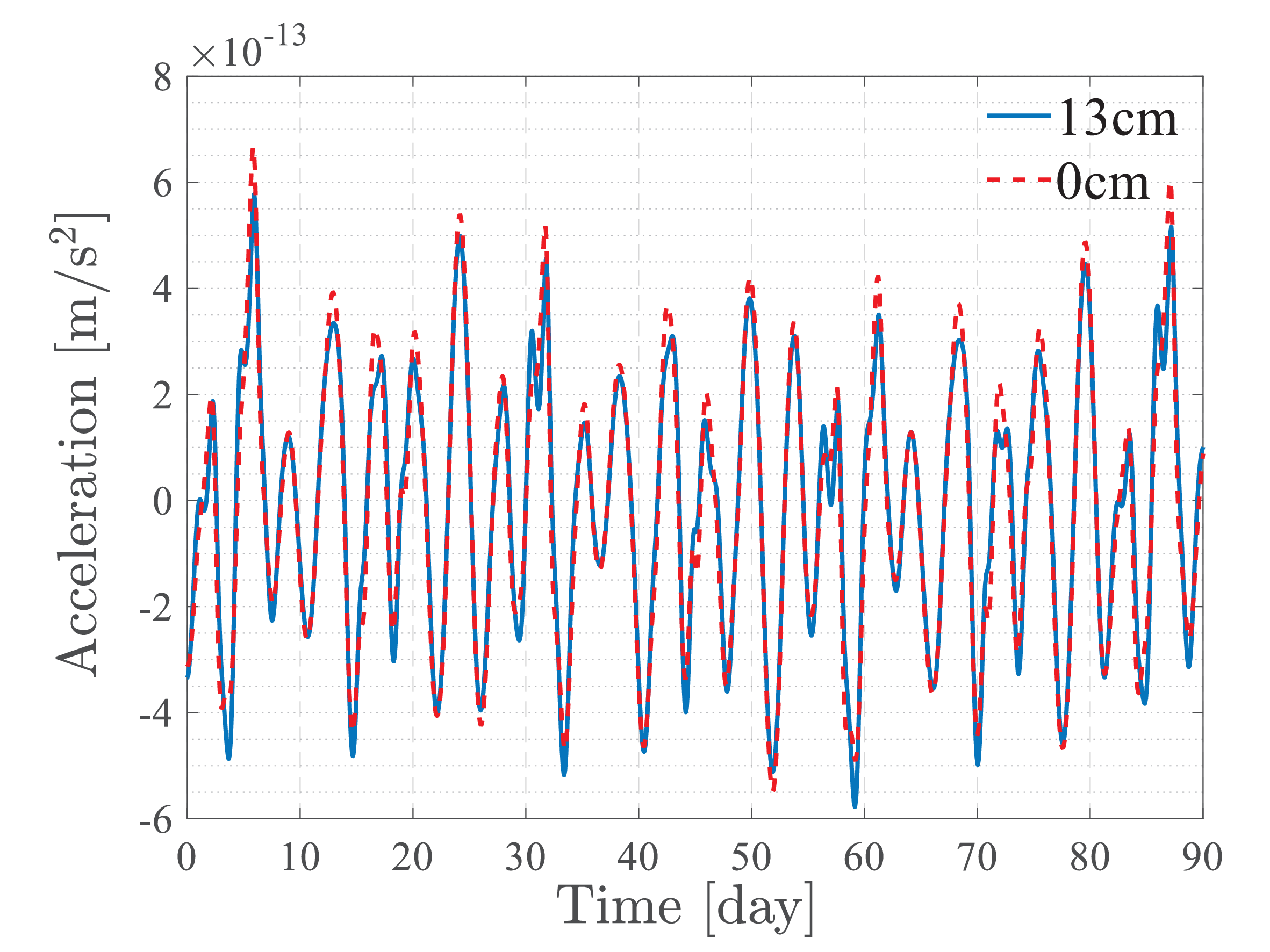}
\caption{\label{fig:pivotDiffAll} The control accelerations of TM$_1$ for different pivot positions (0 cm, 13 cm) with a 40 cm TM separation for three months. }
\end{figure}

To help confirm the results, we can further compare the time evolutions of the electrostatic control accelerations of $\mathrm{TM}_1$ for two different pivot positions with a 40 cm TM separation (see Fig. \ref{fig:pivotDiffAll}). The plot and calculation show that both the maximum absolute value and the root mean square of the control acceleration are greater when the pivot is at the origin than when the pivot is displaced 13 cm forward. It indicates that the inertial accelerations resulting from the pivot deviating from the EH center can help to offset the gravity gradient (see Eq. (\ref{eq:TMforce})). This design is beneficial for counter-balancing the heavy telescope at the front part of the MOSA. 

\section{Nominal Attitude Control of Satellites}\label{sec:AttitudeofSatellite}
This section calculates the nominal control forces and torques that are required to maintain the satellites' drag-free orbits and nominal attitudes. This is done with time-varying SRP on the flat-top sunshields of the satellites, and we omit the effect of MOSA rotation on the satellite dynamics \cite{vidano2020DYN} which is estimated to be negligible. Then, the Kuhn-Tucker algorithm \cite{gordon2012KT} is employed to allocate thrust to individual nozzles. The process is considered successful if positive-value solutions can be found. Moreover, we search for optimized nozzle orientations that can lower the average thrust output, i.e., fuel consumption. Also the assessment is performed for a total period of four months (e.g., 2034/5/24 – 2034/9/22) by adding 15-day margins before and after three-month observation windows. 

\subsection{Estimated total thrusts and torques} \label{sec:thrustreq}
\begin{figure}[ht]
\begin{minipage}{0.48\textwidth}
  \centering
  \includegraphics[width=\textwidth]{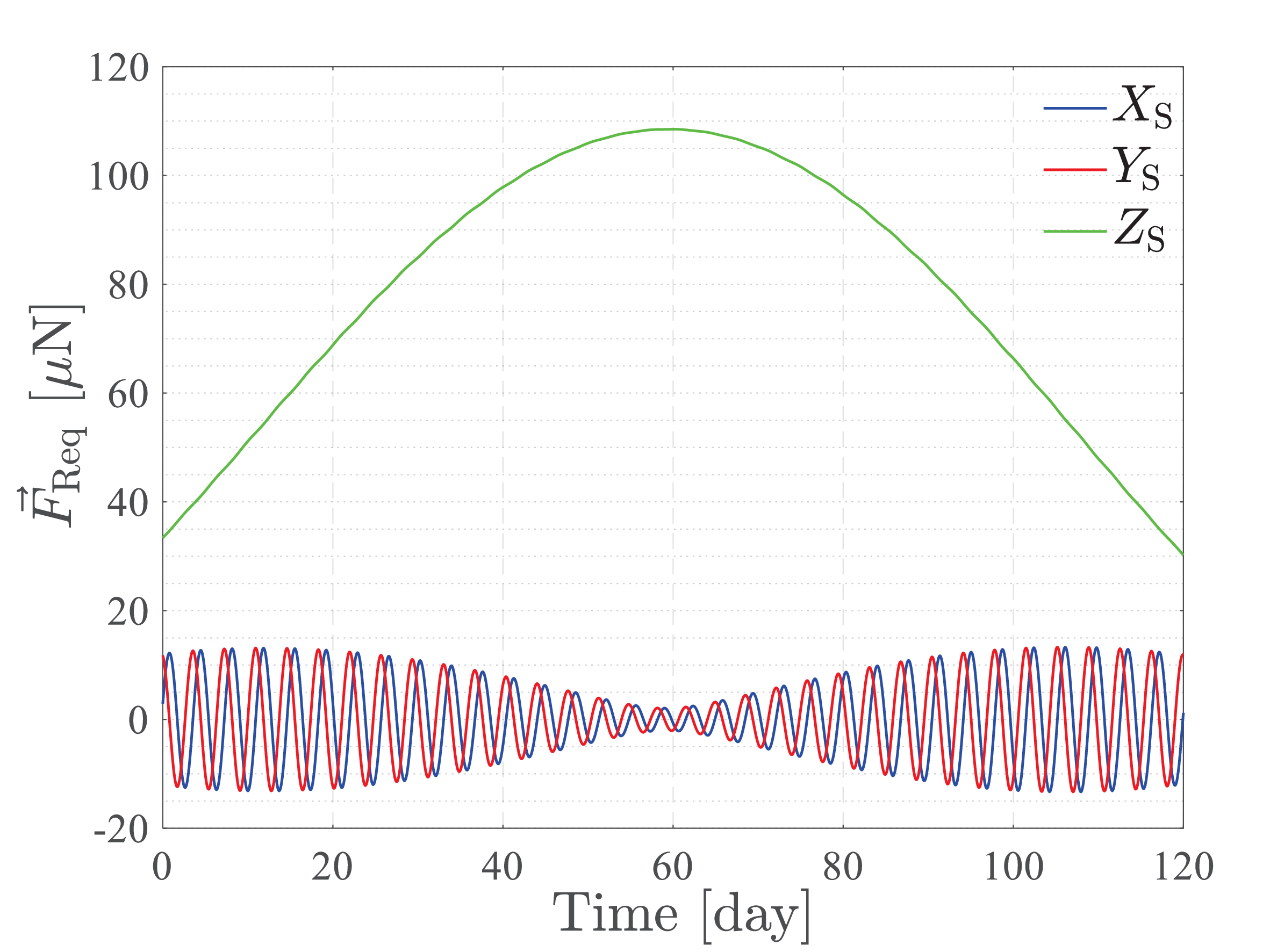}
\end{minipage}
\begin{minipage}{0.48\textwidth}
  \centering
  \includegraphics[width=\textwidth]{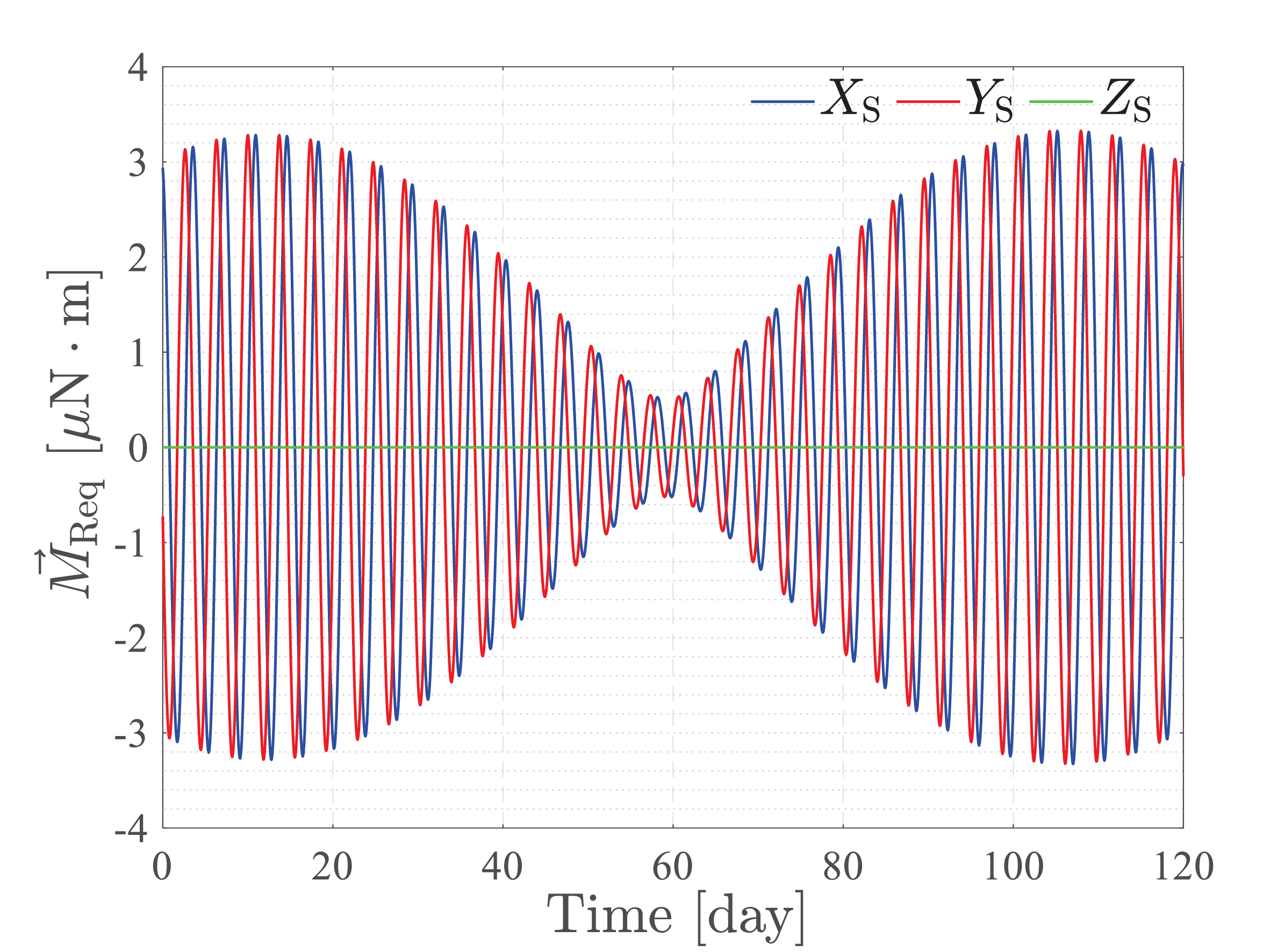}
\end{minipage}
\caption{\label{fig:ThrustReq} Estimated total thrust and torque for one TianQin satellite to maintain the drag-free orbit and nominal attitude. }
\end{figure}

Micro-newton thrusters are used to offset non-gravitational forces and steer the satellites in the desired orbits and attitudes. The main sources of disturbances are the SRP and the thermal radiation emitted by the satellites themselves. The satellites' nominal orbits and attitudes, along with the Sun's position and simulated satellite temperatures (e.g., $\sim 40^\circ$C at the sunshields \cite{wang2023thermal}, see also Table \ref{tab:SCparam}), are combined to estimate the total thrusts and torques required for four months. The typical result is shown for one satellite in Fig. \ref{fig:ThrustReq}. The plots for the other two satellites are quite similar and hence omitted here. 

It can be seen from the plots that except along the $\vec{Z}_\mathrm{S}$-direction, the variations of the thrusts and torques has the same period of the orbit (3.6 days). This is due to the fact that with the telescopes aiming at other satellites, the $\vec{X}_\mathrm{S}$-axis of the satellite is always Earth-pointing and hence the satellite rotates at the same rate with the orbit. In addition, the variations are modulated by the slow-varying solar angle with respect to the constellation plane. 

\subsection{Thrust allocations}
Thruster layout affects thrust allocation among the nozzles. The two preliminary configurations have been given in Sec. \ref{sec:SC_model} and Fig. \ref{fig:ThrustersArrang}. Note that the satellites lack thrusters pointing in the $-\vec{Z}_\mathrm{S}$ direction. Instead, the SRP can be used as a virtual thruster to work jointly with the others \cite{armano2019LPFthrusters}. Under this constraint and thruster configurations, we can use the Kuhn-Tucker optimization algorithm to determine the availability of positive-value solutions for various nozzle orientations.

To represent nozzle orientations, we use pitch and yaw angles defined in a coordinate system where the $x$-axis aligns with the outward normal of the side panel, and the $z$-axis aligns with the satellite's $\vec{Z}_\mathrm{S}$-axis, with $y$-axis to complete the right-hand system (see Fig. \ref{fig:ThrustersArrang}). By rotating around the $y$-axis with the pitch angle and then around $z$-axis with the yaw angle, one obtains the directional vector of one nozzle. The two nozzles of the same cluster are mirror-symmetric with respect to the $x$-$z$ plane. The parameter space falls within the range of $(0^\circ, 90^\circ)$ for both pitch and yaw. 

\begin{figure}[ht]
\centering
\includegraphics[width=0.48\textwidth]{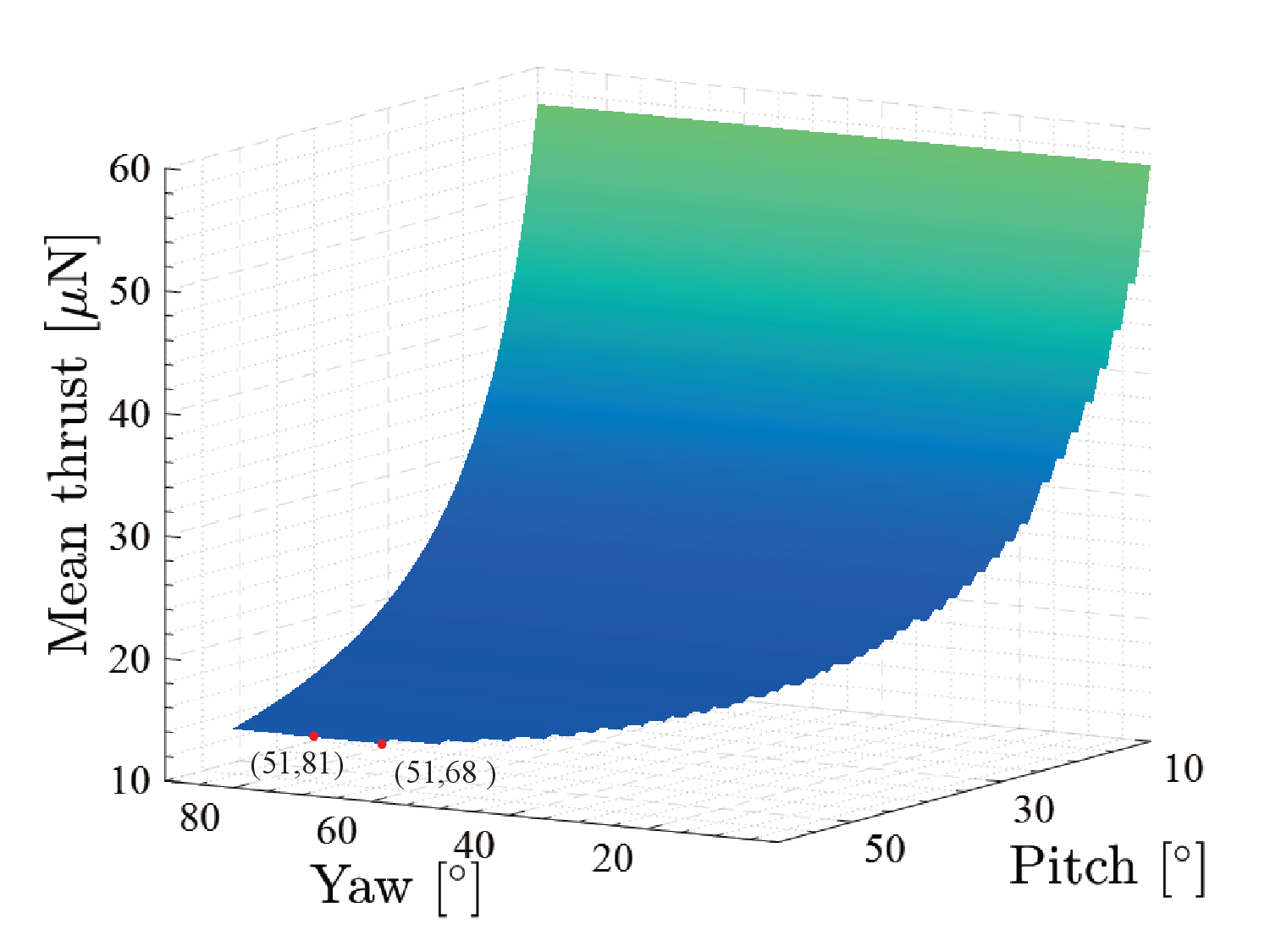}
\caption{\label{fig:ThrustMean} Selectable nozzle orientations for the four-cluster configuration with corresponding mean thrusts of the eight nozzles averaged for four months. The red dots mark optimized nozzle orientations. }
\end{figure}

By exhausting the parameter space with an step size of $1^\circ$, we have identified the selectable range of nozzle angles that can generate positive thruster outputs over four months. The result is shown in Fig. \ref{fig:ThrustMean}, with the vertical scale denoting the average thrust calculated from all eight nozzles.

\begin{figure}[ht]
\centering
\includegraphics[width=0.48\textwidth]{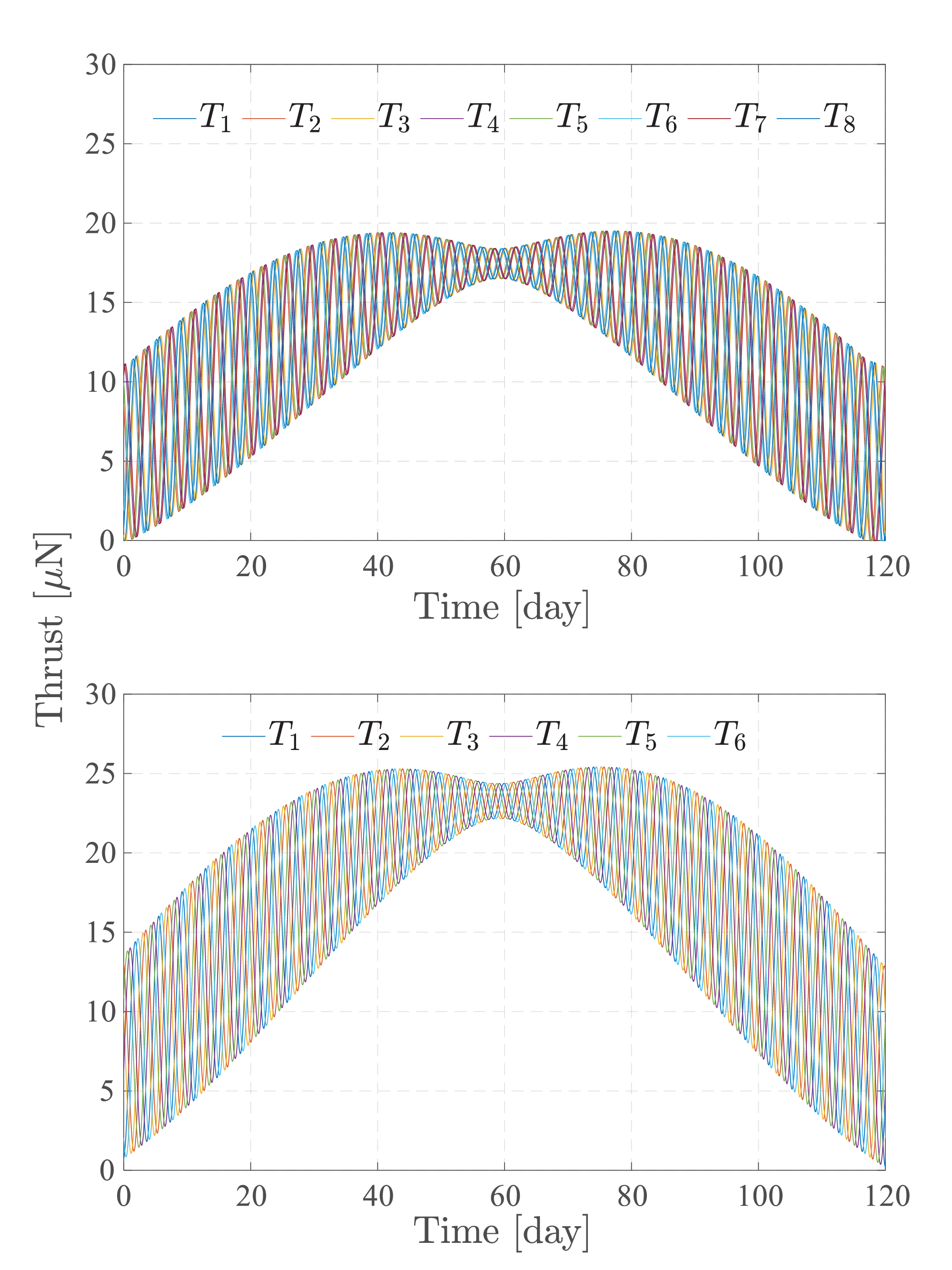}
\caption{\label{fig:Thrusts} The thrust variations over four months, allocated for the four-cluster and three-clusters configurations, and with the optimized nozzle orientation at a 51$^\circ$ pitch and a 81$^\circ$ yaw. }
\end{figure}

According to the result, the nozzle orientation with the least variation and average thrust is obtained at a 51$^\circ$ pitch and a 81$^\circ$ yaw for the four-cluster design. The case of the three-cluster design yields the same optimized orientations. The corresponding thrust variations of the nozzles are shown Fig. \ref{fig:Thrusts}. The plots show that the four-cluster configuration has narrower output ranges and a less average thrust than the three-cluster configuration. Both designs can fulfill the TianQin requirements, which warrants further trade-offs. In addition, to avoid plume impingement, the yaw angle can be relaxed down to 68$^\circ$ without altering the average thrust (see Fig. \ref{fig:ThrustMean}).

\section{Conclusion and Discussion}\label{sec:conclusion}
In this paper, we have assessed the applicability of the two TMs and telescope pointing scheme to the TianQin mission under the geocentric perturbed orbits and orbital gravity gradients, and also optimized certain basic mechanical parameters for a better adaptation. This is done by estimating the required electrostatic control forces and toques on the TMs and comparing them with the allowed maximum values, and by finding thrust allocation solutions for the satellite DFAC under the constraint of the satellite configuration and varying solar angles. The estimations are based on the geometric relation of the orbit-attitude coupling and can work through without the need of detailed control algorithms. Two main conclusions can be drawn here. 

\begin{enumerate}
    \item The required electrostatic control accelerations and angular accelerations for the TM suspension control along the non-sensitive axes are estimated at $\sim\!10^{-13}$ m/s$^2$ and $\sim\!10^{-13}$ rad/s$^{2}$, respectively, which are well below the requirements from the acceleration noise budget. Moreover, their magnitudes and the required total thrust on the satellite can be minimized by configuring the CoMs of the two TMs and the satellite symmetrically in syzygy, and by offsetting the MOSA pivot from the EH center forward along the sensitive axis by $\sim 10$ cm. This pivot offsetting is found to be effective in having the inertial acceleration and the gravity gradient acceleration partially cancel each other. 
    \item Both the three-cluster and four-cluster configurations of the micro-thrusters are capable of sustaining the drag-free orbits and the nominal attitudes of the satellites for consecutively four months of science observation. A combination of a 51$^\circ$ pitch angle and $68-81^\circ$ yaw angles of the thrust direction relative to the installation panel has been identified to yield smallest average thrust and thrust variations. 
\end{enumerate}

As no principle issues are identified, the findings support adopting the two TMs and telescope pointing scheme as the current baseline for TianQin, also given that the scheme has become more mature technologically than other options. The analyses also provide useful reference to the system design of the MOSA and satellite. For future works, the dynamic model (TQDYN) should be further extended to include, e.g., the self-gravity from the satellite, and an integration with high-precision orbit propagation (TQPOP, \cite{zhang2021effect}) and the split interferometry (TQTDI, \cite{zheng2023prd}) is also of great interest to the mission (see, e.g., \cite{heisenberg2023lisa}). 

\begin{acknowledgments}
The authors thank Dexuan Zhang, Bobing Ye, Jihe Wang, Hsien-Chi Yeh, Ze-Bing Zhou, Jun Luo, and the anonymous referees for helpful discussions and comments. X. Z. is supported by the National Key R\&D Program of China (Grant No. 2022YFC2204600 and 2020YFC2201202), NSFC (Grant No. 12373116), and Fundamental Research Funds for the Central Universities, Sun Yat-sen University (Grant No. 23lgcxqt001). 
\end{acknowledgments}

\appendix*

\section{Estimation of orbit deviation} \label{App:orbit_dev}
The deviation from pure-gravity orbits due to DFAC is a factor that needs to be considered in the orbit propagation. LISA has made relevant estimates on the constellation stability \cite{martens2021trajectory, povoleri2006lisa}. To evaluate the magnitude of the deviation for TianQin, we use the following equation describing the relative motion between the drag-free controlled satellite and an ideally free-falling satellite: 
\begin{equation}
\begin{aligned}
^S\ddot{\vec{R}}_{SC} \approx\, &{^{S}{\vec{A}}_\mathrm{c,S}}+\Gamma_{S} \: {^{S}\!{\vec{R}}_\mathrm{SC}}-{^{S}\vec{\omega}_\mathrm{S}} \times ({^{S}\vec{\omega}_\mathrm{S}} \times{^{S}\!\vec{R}_\mathrm{SC}})\\
&-2{^{S}\!\vec{\omega}_\mathrm{S}}\times{^{S}\!\dot{\vec{R}}_\mathrm{SC}}-{^{S}\dot{\vec{\omega}}_\mathrm{S}}\times{^{S}\!{\vec{R}}_\mathrm{SC}}.
\end{aligned}
\end{equation}

It approximates the differential gravitational acceleration between the actual satellite and its ideal position by using the gravity gradient $\Gamma_{S}$ at the latter position. The terms $^S{\vec{R}}_\mathrm{SC}$, $^S\dot{\vec{R}}_\mathrm{SC}$, $^S\ddot{\vec{R}}_\mathrm{SC}$ describe the relative state of the actual satellite from its ideal state. The term ${^{S}{\vec{A}}_\mathrm{c,S}}$ is the external acceleration provided by thrusters, and equal to $^{S}\vec{G}$ in the Eq. (\ref{eq:G}), and the next three terms corresponding to the inertial acceleration caused by the moving $\mathcal{S}$ frame. 

\begin{figure}[htb]
\centering
\includegraphics[width=0.48\textwidth]{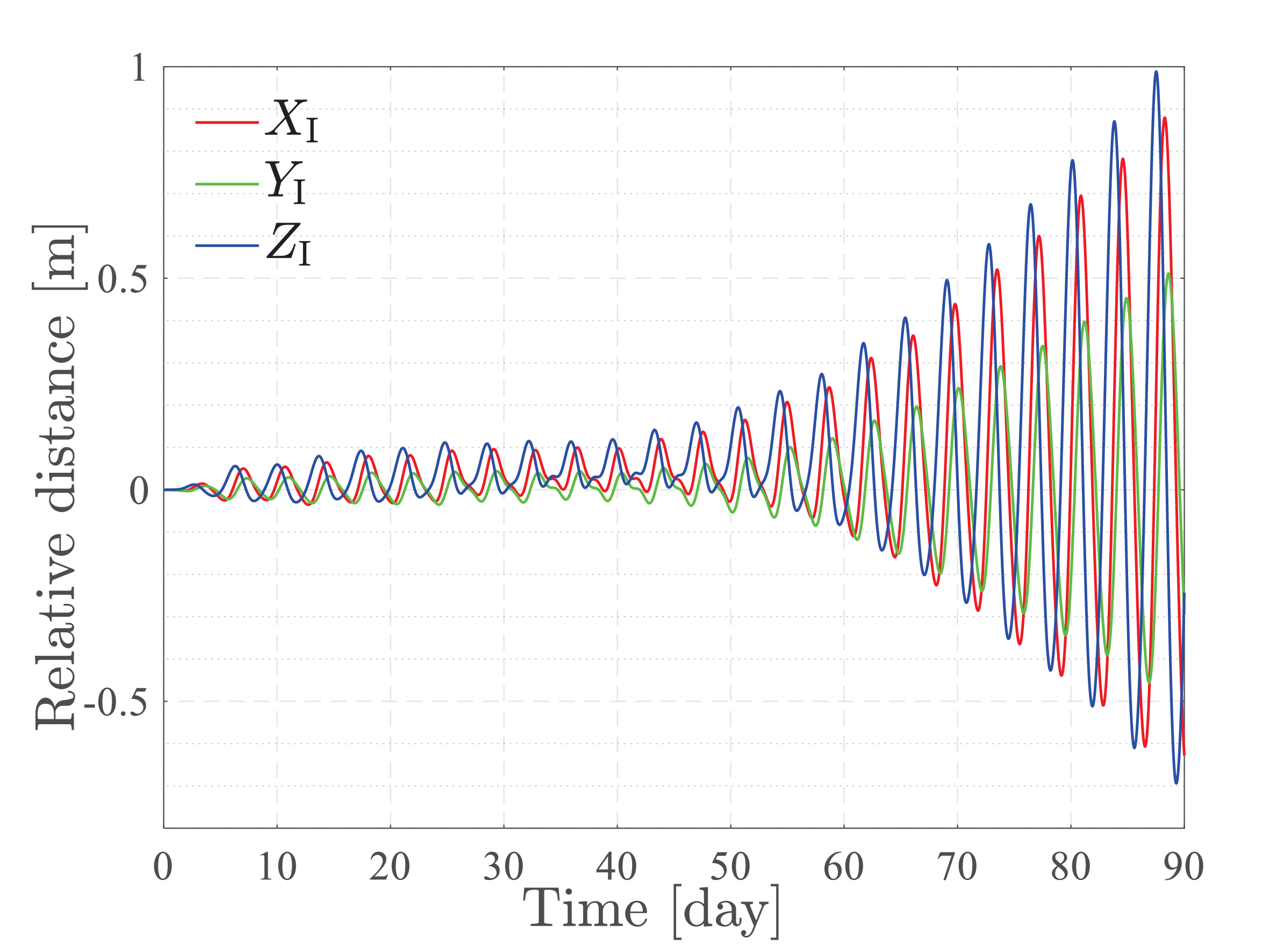}
\caption{\label{fig:RM}The relative distance between the drag-free controlled orbit and the ideally free-falling orbit. The red, green and blue lines represent the $x$, $y$, and $z$-components of the relative position vector in the $\mathcal{I}$ frame.}
\end{figure}

The result is shown in Fig. \ref{fig:RM}. The satellite's deviation from its ideally free-falling orbit over a three-month period is in the order of meters. Although the real deviation may be larger due to self-gravity, this deviation is negligible in regard of the constellation stability and nominal attitude variations of the satellites. 

\nocite{*}
\bibliography{bibliography}
\end{document}